\documentclass[12pt,prd,onecolumn,showpacs,amsmath,amssymb,aps,floats,floatfix,nofootinbib]{revtex4-1}

\usepackage[colorlinks=true,urlcolor=blue,anchorcolor=blue,citecolor=blue,filecolor=blue,linkcolor=blue,menucolor=blue,pagecolor=blue,linktocpage=true]{hyperref} 

\usepackage[inline]{enumitem}
\usepackage[multidot]{grffile} 
\usepackage{dcolumn}
\usepackage{bm}
\usepackage{amsmath}
\usepackage{amsfonts}
\usepackage{amssymb}
\usepackage{color}
\usepackage{latexsym}
\usepackage{slashed} 
\usepackage{pstricks}
\usepackage{indentfirst}
\usepackage{mathrsfs}
\usepackage{multirow}
\usepackage{epsfig,psfrag}
\usepackage{subfigure}
\usepackage{mathtools}
\usepackage{setspace} 
\usepackage[utf8]{inputenc} 
\usepackage[scientific-notation=true]{siunitx} 
\usepackage{verbatim}

\graphicspath{{fig/}}

\newcommand{\SUtwoL}{\mathrm{SU}(2)_\mathrm{L}}
\newcommand{\UoneY}{\mathrm{U}(1)_\mathrm{Y}}
\newcommand{\Uone}{\mathrm{U}(1)}
\newcommand{\svann}{\left<\sigma_\mathrm{ann}v\right>}

\makeatother

\allowdisplaybreaks 

\begin{document}

\title{Phase transition gravitational waves from pseudo-Nambu-Goldstone dark matter and two Higgs doublets}
\author{Zhao Zhang}
\author{Chengfeng Cai}
\author{Xue-Min Jiang}
\author{Yi-Lei~Tang}\email[Corresponding author. ]{tangylei@mail.sysu.edu.cn}
\author{Zhao-Huan Yu}\email[Corresponding author. ]{yuzhaoh5@mail.sysu.edu.cn}
\author{Hong-Hao Zhang}\email[Corresponding author. ]{zhh98@mail.sysu.edu.cn}
\affiliation{School of Physics, Sun Yat-Sen University, Guangzhou 510275, China}

\begin{abstract}
We investigate the potential stochastic gravitational waves from first-order electroweak phase transitions in a model with pseudo-Nambu-Goldstone dark matter and two Higgs doublets. The dark matter candidate can naturally evade direct detection bounds, and can achieve the observed relic abundance via the thermal mechanism. Three scalar fields in the model obtain vacuum expectation values, related to phase transitions at the early Universe. We search for the parameter points that can cause first-order phase transitions, taking into account the existed experimental constraints.
The resulting gravitational wave spectra are further evaluated.
Some parameter points are found to induce strong gravitational wave signals, which have the opportunity to be detected in future space-based interferometer experiments LISA, Taiji, and TianQin.
\end{abstract}

\maketitle
\tableofcontents

\section{Introduction}

It is conventionally believed that dark matter (DM) originates from thermal production at the early Universe~\cite{Bertone:2004pz,Feng:2010gw,Young:2016ala}.
Thus, the DM relic abundance would be determined by the annihilation cross section at the freeze-out epoch.
The relic abundance observation suggests that the natural strength of the DM couplings to standard model (SM) particles should be close to the weak interaction strength. This motivates the worldwide establishment of various direct detection experiments searching for nuclear recoil signals induced by DM scattering.
Nonetheless, no DM signal is robustly found in these experiments so far, leading to stringent constraints on the DM-nucleon scattering cross section~\cite{Akerib:2016vxi,Cui:2017nnn,Aprile:2018dbl}.
Therefore, the thermal production paradigm faces a serious challenge.

Such a situation can be circumvented if one can effectively suppress DM-nucleon scattering at zero momentum transfer without reducing DM annihilation at the freeze-out epoch.
An appealing approach to achieve this is provided by Higgs-portal pseudo-Nambu-Goldstone boson (pNGB) DM models~\cite{Gross:2017dan,Azevedo:2018exj,Ishiwata:2018sdi,Huitu:2018gbc,Alanne:2018zjm,Kannike:2019wsn,Karamitros:2019ewv,Cline:2019okt,Jiang:2019soj,Arina:2019tib,Abe:2020iph,Okada:2020zxo,Glaus:2020ihj,Abe:2020ldj}, where the DM candidate is a pNGB protected by a global symmetry which is softly broken by quadratic mass terms.
The pNGB nature makes the tree-level DM-nucleon scattering amplitude vanish in the zero momentum transfer limit~\cite{Gross:2017dan}.
Although loop corrections give rise to a nonzero scattering cross section, one-loop calculations show that near future direct detection experiments would not be able to probe the pNGB DM~\cite{Azevedo:2018exj,Ishiwata:2018sdi,Glaus:2020ihj}.
Therefore, other experimental approaches are crucial for exploring these models.

There are various ways to experimentally test DM models, including direct and indirect DM detection, collider searches, etc. The discovery of gravitational waves (GWs) by LIGO~\cite{Abbott:2016blz} provides a new path. DM fields could be relevant to strong first-order electroweak phase transitions (EWPTs) that produce detectable stochastic GW signals in the proposed future GW experiments~\cite{Jiang:2015cwa, Chala:2016ykx,Chao:2017vrq,Beniwal:2017eik,Huang:2017rzf,Huang:2017kzu,Hektor:2018esx,Baldes:2018emh,Madge:2018gfl,Beniwal:2018hyi,Bian:2018mkl,Bian:2018bxr,Shajiee:2018jdq,Mohamadnejad:2019vzg,Kannike:2019mzk,Paul:2019pgt,Chen:2019ebq,Barman:2019oda,Chiang:2019oms,Borah:2020wut,Kang:2020jeg,Pandey:2020hoq,Han:2020ekm,Alanne:2020jwx,Wang:2020wrk,Ghosh:2020ipy,Huang:2020mso,Chao:2020adk}.
Such stochastic GWs typically peak around the mHz frequency band~\cite{Mazumdar:2018dfl}, to which ground-based laser interferometers are not sensitive.
Nonetheless, future space-based GW interferometer plans, e.g., LISA~\cite{Audley:2017drz}, TianQin~\cite{Luo:2015ght,Hu:2018yqb,Mei:2020lrl}, Taiji~\cite{Hu:2017mde,Guo:2018npi}, DECIGO~\cite{Seto:2001qf,Kudoh:2005as}, and BBO~\cite{Ungarelli:2005qb,Cutler:2005qq}, are able to probe sub-Hz bands and look for the stochastic GW signals.

The minimal setup of the pNGB DM involves a complex scalar singlet with a global $\Uone$ symmetry and a quadratic term that softly breaks $\Uone$ into $Z_2$~\cite{Gross:2017dan}.
The singlet and the SM Higgs doublet together could induce two-step phase transitions.
Nevertheless, a study~\cite{Kannike:2019wsn} showed that such phase transitions can only be of second order and impossible to produce stochastic GWs.
Further studies tried to introduce extra terms to break the $\Uone$ symmetry, e.g., the soft cubic terms~\cite{Kannike:2019mzk}, or the most general breaking terms~\cite{Alanne:2020jwx}.
These efforts successfully achieved first-order phase transitions (FOPTs) and stochastic GWs, but the essential merit of the vanishing tree-level DM-nucleon scattering in the zero momentum transfer limit is sacrificed.

It would be rather interesting if we can find out a pNGB DM setup that allows both the vanishing DM-nucleon scattering and detectable stochastic GW signals.
Inspired by the notable GW signals from the strong FOPTs obtained in the two-Higgs-doublet models~\cite{Dorsch:2016nrg,Wang:2019pet,Zhou:2020xqi},  we study the possibility of extending the minimal pNGB DM setup with an additional Higgs doublet~\cite{Jiang:2019soj}, which is expected to involve more phase transition patterns. The corresponding phase transitions and stochastic GWs will be studied in this paper in detail.

In the following Sec.~\ref{sec:model}, we briefly introduce the model and the particle masses.
Existed experimental bounds are described in Sec.~\ref{sec:bound}.
The effective potential at finite temperature is constructed in Sec.~\ref{sec:eff_poten}.
We analyze key properties of the EWPT in Sec.~\ref{sec:pts}, which are relevant to the GW spectra discussed in Sec.~\ref{sec:gws}.
Sec.~\ref{sec:num} give numerical analyses of GW signals based on random parameter scans.
We summarize the paper in Sec.~\ref{sec:concl}.

\section{The model}
\label{sec:model}

In this section, we briefly describe the model we are interested in.
More details can be found in Ref.~\cite{Jiang:2019soj}.
This model involves two $\SUtwoL$ Higgs doublets $\Phi_1$ and $\Phi_2$, both carrying hypercharge $1/2$, and a complex scalar $S$ which is a SM gauge singlet.
The Lagrangian respects a global $\Uone$ symmetry $S\to e^{i\alpha}S$ explicitly violated into a $Z_2$ symmetry $S\to -S$ by soft breaking quadratic terms.
The $\Uone$ symmetry is further spontaneously broken after $S$ develops a vacuum expectation value (VEV).
Then the imaginary part of $S$ becomes a pNGB, acting as a DM candidate.

As in the simplified versions of the two-Higgs-doublet models~\cite{Branco:2011iw}, we assume that the scalar potential respects a $Z_2$ symmetry $\Phi_1 \to -\Phi_1$ or $\Phi_2 \to -\Phi_2$ which is only softly broken by quadratic terms.
Moreover, $CP$ conservation is assumed in the scalar sector, leading to only real coefficients.
The potential satisfying these two assumptions and the global $\Uone$ symmetry reads
\begin{align}\label{eq:V_Phi,S}
V_{\Phi_i,S} &= m_{11}^2|{\Phi_1}|^2 + m_{22}^2|{\Phi_2}|^2 - m_{12}^2(\Phi _1^\dag {\Phi_2} + \Phi_2^\dag {\Phi_1}) + \frac{\lambda_1}{2}|{\Phi_1}{|^4} + \frac{\lambda_2}{2}|{\Phi_2}{|^4}
\notag\\
&\quad + {\lambda _3}|{\Phi _1}|^2|{\Phi _2}|^2 + {\lambda _4}|\Phi _1^\dag {\Phi _2}|^2 + \frac{{{\lambda _5}}}{2}[{(\Phi _1^\dag {\Phi _2})^2} + {(\Phi _2^\dag {\Phi _1})^2}] \notag\\
&\quad - m_S^2|S|^2 + \frac{{{\lambda _S}}}{2}|S{|^4} + {\kappa _1}|{\Phi _1}|^2|S|^2 + {\kappa _2}|{\Phi _2}|^2|S|^2.
\end{align}
The $\Uone$ soft breaking terms
\begin{equation}\label{eq:V_soft}
V_{\mathrm{soft}} =  - \frac{m'^2_S}{4}S^2 + \mathrm{H.c.}
\end{equation}
are further introduced in the potential.
Thus, the total scalar potential is $V = V_{\Phi_i,S} + V_\mathrm{soft}$.

We can always make the soft breaking parameter $m'^2_S$ real and positive through a phase redefinition of $S$.
Consequently, the potential respects a dark $CP$ symmetry $S\to S^*$.
For $m'^2_S > 0$, the VEV of $S$ developed must be real, and the dark $CP$ symmetry remain unbroken, ensuring that the imaginary part of $S$ acts as a stable DM candidate~\cite{Gross:2017dan}.

If the charged component of $\Phi_1$ or $\Phi_2$ gains a nonzero VEV, the photon would become massive, and the theory is unacceptable.
If the neutral component of $\Phi_1$ or $\Phi_2$ develops an imaginary VEV, $CP$ would be spontaneously broken.
Detailed discussions on vacuum configurations and parameter relations in general two-Higgs-doublet models can be found in Ref.~\cite{Ginzburg:2004vp}.
Here we are particularly interested in the case that only the neutral real parts of $\Phi_1$, $\Phi_2$, and $S$ develop nonzero VEVs $v_1$, $v_2$, and $v_s$, respectively.
Thus, at the zero temperature, these scalar fields can be expanded as
\begin{eqnarray}
\Phi_1 &=& \begin{pmatrix}
{\phi_1^+ }  \\
{(v_1 + {\rho_1} + i{\eta _1})/\sqrt 2 }  \\
\end{pmatrix},
\\
\Phi_2 &=& \begin{pmatrix}
{\phi_2^+ }  \\
{(v_2 + {\rho_2} + i{\eta_2})/\sqrt 2 }  \\
\end{pmatrix},
\\
S &=& \frac{v_s + s + i\chi }{\sqrt 2 }.
\end{eqnarray}
The potential is minimized at $(v_1, v_2, v_s)$, leading to three stationary point conditions,
\begin{eqnarray}
m_{11}^2 &=& {m}_{12}^2 \tan\beta - \frac{1}{2}({\lambda _1}v_1^2 + \lambda_{345}v_2^2 + {\kappa _1}v_s^2), \label{eq:stationary:1}\\
m_{22}^2 &=& {m}_{12}^2 \cot\beta - \frac{1}{2}( {\lambda _2}v_2^2 +\lambda_{345}v_1^2 +{\kappa _2}v_s^2),\label{eq:stationary:2} \\
m_S^2 &=& \frac{1}{2}({\kappa _1}v_1^2 + {\kappa _2}v_2^2 + {\lambda _S}v_s^2 - m'^2_S),\label{eq:stationary:3}
\end{eqnarray}
where
\begin{eqnarray}
\beta &\equiv& \arctan \frac{v_2}{v_1},
\\
\lambda_{345} &\equiv& {\lambda_3} + {\lambda_4} + {\lambda_5}.
\end{eqnarray}

The VEVs contribute to a $3\times 3$ mass-squared matrix for the $CP$-even neutral scalars $\rho_1$, $\rho_2$, and $s$.
The eigenvalues $m_{h_1}^2$, $m_{h_2}^2$, and $m_{h_3}^2$ of this matrix are the masses squared for the mass eigenstates $h_1$, $h_2$, and $h_3$, respectively.
One of $h_i$ must behave as a SM-like Higgs boson with a mass of $\sim 125~\si{GeV}$, satisfying the experimental observations.
After rotations with the angle $\beta$, the $CP$-odd neutral scalars $\eta_1$ and $\eta_2$ are transformed into the mass eigenstates $G^0$ and $a$, while the charged scalars $\phi_1^+$ and $\phi_2^+$ are transformed into the mass eigenstates $G^+$ and $H^+$.
$G^0$ and $G^\pm$ are the Nambu-Goldstone bosons eaten by the $Z$ and $W^\pm$ gauge bosons.
$a$ and $H^\pm$ are extra Higgs bosons, whose masses squared are given by
\begin{eqnarray}
m_a^2 &=& \frac{1}{\sin\beta \cos\beta} \left(m_{12}^2 - \lambda_5 v_1 v_2\right),
\\
m_{H^\pm}^2 &=& \frac{1}{\sin\beta \cos\beta} \left[m_{12}^2 - \frac{1}{2}(\lambda_4 + \lambda_5) v_1 v_2\right].
\end{eqnarray}
For $m'^2_S = 0$, the neutral boson $\chi$ is a massless  Nambu-Goldstone boson due to the global $\Uone$ symmetry. The soft breaking terms endow the pNGB $\chi$ with a mass of
\begin{equation}
m_\chi = m'_S.
\end{equation}
Besides, $\chi$ only appears in pairs in the interaction terms, guaranteeing its stability to become a DM candidate. The pNGB feature also eliminates the tree-level $\chi$-nucleon scattering amplitude in the zero momentum transfer limit without any parameter tuning~\cite{Gross:2017dan,Jiang:2019soj}.
Thus, this model is hardly constrained by DM direct detection experiments.

The masses of the $W$ and $Z$ gauge bosons are given by
\begin{equation}
m_W = \frac{g v}{2},\quad
m_Z = \frac{v}{2}\sqrt{g^2+g'^2},
\end{equation}
where $v \equiv\sqrt{v_1^2 + v_2^2}$, and $g$ and $g'$ denote the $\SUtwoL$ and $\UoneY$ gauge couplings, respectively.
Thus, we observe that $v$ is equivalent to the Higgs VEV in the SM and can be expressed as $v = (\sqrt{2} G_\mathrm{F})^{-1/2}$, where $G_\mathrm{F}$ is the Fermi constant.

For the two Higgs doublets, four types of Yukawa couplings  without tree-level flavor-changing neutral currents (FCNCs) can be constructed~\cite{Glashow:1976nt,Paschos:1976ay,Branco:2011iw}.
In this paper, we only focus on the type-I and type-II Yukawa couplings, whose  Lagrangians are respectively given by
\begin{eqnarray}
{\mathcal{L}_{{\mathrm{Y,I}}}} &=&  - \tilde y_d^{ij}{{\bar Q}_{i{\mathrm{L}}}}{d'_{j{\mathrm{R}}}}{\Phi _2} - \tilde y_u^{ij}{{\bar Q}_{i{\mathrm{L}}}}{u'_{j{\mathrm{R}}}}{{\tilde \Phi }_2} - {\tilde y_{{\ell _i}}}{{\bar L}_{i{\mathrm{L}}}}{\ell _{i{\mathrm{R}}}}{\Phi _2} + \mathrm{H.c.},\\
{\mathcal{L}_{{\mathrm{Y,II}}}} &=&  - \tilde y_d^{ij}{{\bar Q}_{i{\mathrm{L}}}}{d'_{j{\mathrm{R}}}}{\Phi _1} - \tilde y_u^{ij}{{\bar Q}_{i{\mathrm{L}}}}{u'_{j{\mathrm{R}}}}{{\tilde \Phi }_2} - {\tilde y_{{\ell _i}}}{{\bar L}_{i{\mathrm{L}}}}{\ell _{i{\mathrm{R}}}}{\Phi _1} + \mathrm{H.c.},
\end{eqnarray}
where $L_{i\mathrm{L}} \equiv (\nu_{i\mathrm{L}}, \ell_{i\mathrm{L}})^\mathrm{T}$, $Q_{i\mathrm{L}} \equiv (u'_{i\mathrm{L}}, d'_{i\mathrm{L}})^\mathrm{T}$, ${\tilde \Phi }_2\equiv i\sigma^2 \Phi_2^*$, and $i,j=1,2,3$.
The Yukawa coupling matrices $\tilde y_d^{ij}$ and $\tilde y_u^{ij}$ can be diagonalized by unitary matrices, which transform the gauge eigenstates $u'_i$ and $d'_i$ into the mass eigenstates $u_i$ and $d_i$.
We remark that due to the similarity of the Yukawa couplings in the quark sectors and the smallness of the ones in the leptonic sectors, many of the following analyses for the type-I (type-II) case can be cast to the lepton specific (flipped) case.

\section{Experimental bounds}
\label{sec:bound}

In our analyses, we carry out random scans in the parameter space.
The following 12 parameters are adopted as the free parameters:
\begin{equation}
\lambda_1,~~ \lambda_2,~~ \lambda_3,~~ \lambda_4,~~ \lambda_5,~~ \lambda_S,~~ \kappa_1,~~ \kappa_2,~~ \tan\beta,~~ m_\chi,~~ v_s,~~  m_{12}^2.
\end{equation}
Each parameter point in the scans should be tested by existed experimental bounds.

Firstly, we require that $m_{h_i}^2$ ($i=1,2,3$), $m_a^2$, and $m_{H^+}^2$ should be positive to guarantee physical scalar masses.
Moreover, in order to ensure that the scalar potential is bounded from below, the following conditions from copositivity criteria~\cite{Klimenko:1984qx,Kannike:2012pe} should be satisfied:
\begin{gather}
\lambda_1 \ge 0, \quad \lambda_2 \ge 0, \quad \lambda_S \ge 0, \label{eq:positive_lambda_12S}\\
a_{12} \equiv \lambda_3 + \sqrt{\lambda_1 \lambda_2} \ge 0, \quad a'_{12} \equiv \lambda_3 + \lambda_4 -|\lambda_5| +\sqrt{\lambda_1 \lambda_2} \ge 0,\\
\quad a_{13} \equiv \kappa_1 + \sqrt{\lambda_1 \lambda_S} \ge 0, \quad a_{23} \equiv \kappa_2 + \sqrt{\lambda_2 \lambda_S} \ge 0,\\
\sqrt{\lambda_1 \lambda_2 \lambda_S} + \lambda_3 \sqrt{\lambda_S} + \kappa_1 \sqrt{\lambda_2} + \kappa_2 \sqrt{\lambda_1} +  \sqrt{2 a_{12} a_{13} a_{23}} \ge 0, \\
\sqrt{\lambda_1 \lambda_2 \lambda_S} + (\lambda_3 + \lambda_4 -|\lambda_5|) \sqrt{\lambda_S} + \kappa_1 \sqrt{\lambda_2} + \kappa_2 \sqrt{\lambda_1} +  \sqrt{2 a'_{12} a_{13} a_{23}} \ge 0.
\end{gather}

Furthermore, we require one of $h_i$ acting as the SM-like Higgs boson with a mass within the $3\sigma$ range of the measured value $m_h = 125.18\pm 0.16~\si{GeV}$~\cite{Tanabashi:2018oca}.
The numerical tool \texttt{Lilith~2}~\cite{Bernon:2015hsa,Kraml:2019sis} is used to test whether the SM-like Higgs boson is consistent with LHC run~1 and run~2 Higgs measurements from ATLAS and CMS.
Parameter points excluded by the data at $95\%$ confidence level (C.L.) are abandoned.

Although FCNCs have been forbidden at tree level, they can arise from loop corrections.
In particular, the loops involving the charged Higgs boson $H^\pm$ significantly contribute to the FCNC $B$-meson decays, depending on $m_{H^\pm}$ and $\tan\beta$.
The analysis by the Gfitter Group~\cite{Haller:2018nnx} shows that the strongest constraint on the type-I (type-II) Yukawa couplings comes from the measurement of the FCNC decay $B_d \to \mu^+ \mu^-$ ($B_s \to \mu^+ \mu^-$ and $B \to X_s \gamma $).
We further reject the parameter points that are excluded  at 95\% C.L. by these flavor physics constraints.

Then we impose the constraints from DM phenomenology.
We utilize \texttt{FeynRules~2}~\cite{Alloul:2013bka} and the \texttt{MadGraph5\_aMC@NLO}~\cite{Alwall:2014hca} plugin \texttt{MadDM 3}~\cite{Ambrogi:2018jqj} to calculate the prediction of the DM relic abundance.
The observed value of the relic abundance from the Planck experiment is given by $\Omega_{\mathrm{DM}}h^2 = 0.1200 \pm 0.0012$~\cite{Aghanim:2018eyx},
where $\Omega_{\mathrm{DM}}$ is the ratio of the DM energy density to the critical density of the Universe and $h$ is the Hubble constant in unit of $100~\si{km~s^{-1}~Mpc^{-1}}$.
Only the parameter points predicting the observed relic abundance are preserved.
\texttt{MadDM 3} is also used to compute the DM annihilation cross section $\svann_\mathrm{d}$ with an average velocity of $\num{2e-5}$, which is corresponding to DM annihilation processes at dwarf spheroidal galaxies.
The 95\% C.L. upper limits on $\svann$ in the $b\bar{b}$ channel from the $\gamma$-ray observations of dwarf galaxies by the Fermi-LAT satellite experiment and the MAGIC Cherenkov telescopes~\cite{Ahnen:2016qkx} are employed to test the parameter points.

Below, we study the effective potential, cosmological phase transitions, and gravitational waves for the parameter points surviving from all the experimental bounds above.

\section{Effective potential}
\label{sec:eff_poten}

In order to investigate the cosmological phase transitions in the model, we need to construct the effective potential.
We assume that only the $CP$-even neutral scalar fields $\rho_1$, $\rho_2$, and $s$ can develop VEVs in the cosmological history.
The effective potential is then expressed as a function of the classical background fields $\tilde \rho_1$, $\tilde \rho_2$, and $\tilde s$.

The tree-level effective potential in terms of the classical fields derived from Eqs.~\eqref{eq:V_Phi,S} and \eqref{eq:V_soft} is
\begin{align}
V_{\mathrm{0}}(\tilde \rho_1, \tilde \rho_2, \tilde s) &= \frac{m_{11}^2}{2}\tilde \rho_1^2 + \frac{m_{22}^2}
{2}\tilde \rho_2^2 - \frac{2m_S^2 + m'^2_S}{4}\tilde s^2 - m_{12}^2\tilde \rho_1 \tilde \rho_2 \notag\\
&\quad + \frac{\lambda_1}{8}\tilde \rho_1^4 + \frac{\lambda_2}
{8}\tilde \rho_2^4 + \frac{\lambda_S}{8}\tilde s^4 + \frac{\lambda_{345}}
{4}\tilde \rho_1^2\tilde \rho_2^2 + \frac{\kappa_1}{4}\tilde \rho_1^2 \tilde s^2 + \frac{\kappa_2}{4}\tilde \rho_2^2\tilde s^2.
\end{align}
Here, $m_{11}^2$, $m_{22}^2$, and $m_{S}^2$ should be expressed as in Eqs.~\eqref{eq:stationary:1}, \eqref{eq:stationary:2}, and \eqref{eq:stationary:3}, respectively.

At zero temperature, the one-loop effective potential $V_1$ receives the Coleman-Weinberg terms~\cite{Coleman:1973jx} in the $\overline{\mathrm{MS}}$ renormalization scheme~\cite{Quiros:1999jp},
\begin{equation}\label{eq:CW}
V_1(\tilde \rho_1,\tilde \rho_2,\tilde s) = \frac{1}{64\pi^2}\sum\limits_i {{n_i}\tilde m_i^4\left({\ln \frac{\tilde m_i^2}{\mu ^2} - C_i} \right)},
\end{equation}
where the sum runs over all the particles $i$ coupling to the classical fields, and $\tilde m_i^2$ are the corresponding particle masses squared in terms of the classical fields.
For the SM fermions, we only take into account the top and bottom quark contributions, and neglect all the other much smaller Yukawa couplings.
Hence, all the particles we include in the calculations are
\begin{equation}
h_1,~h_2,~h_3,~a,~H^\pm,~G^0,~G^\pm,~\chi,~W^\pm,~Z,~\gamma,~t,~b.
\end{equation}
Although the photon $\gamma$ would not contribute to Eq.~\eqref{eq:CW}, its longitudinal mode can contribute to the daisy potential $V_\mathrm{D}$, which will be discussed below.
$\mu$ is the renormalization scale.
For transverse gauge bosons, $C_i = 1/2$, while for longitudinal gauge bosons, scalar bosons and fermions, $C_i = 3/2$.
$n_i$ count the degrees of freedom of the particles, given by
\begin{eqnarray}
n_{h_i} &=& n_{a} = n_{G^0} = n_\chi = n_{Z_\mathrm{L}}= n_{\gamma_\mathrm{L}} = 1,
\\
\quad n_{H^\pm} &=& n_{G^\pm} = n_{Z_\mathrm{T}} = n_{\gamma_\mathrm{T}} = n_{W^\pm_\mathrm{L}} = 2,
\\
n_{W^\pm_\mathrm{T}} &=& 4,\quad  n_t = n_b = -12,
\end{eqnarray}
where the minus signs for $n_t$ and $n_b$ characterize the feature of fermion loops.
The subscripts L and T denote the longitudinal and transverse polarizations of the gauge bosons.

In terms of the classical background fields $\tilde \rho_1$, $\tilde \rho_2$, and $\tilde s$, the elements of the symmetric mass-squared matrix $\tilde M_h^2$ for the $CP$-even neutral scalar bosons are derived as
\begin{eqnarray}
\tilde M_{h,11}^2 &=& m_{11}^2 + \frac{1}{2}(3\lambda_1\tilde \rho_1^2
     + \lambda_{345}\tilde \rho_2^2 + \kappa_1 \tilde s^2),\\
\tilde M_{h,22}^2 &=& m_{22}^2 + \frac{1}{2}(\lambda_{345} \tilde \rho_1^2  + 3\lambda_2 \tilde \rho_2^2 + \kappa_2 \tilde s^2),\\
\tilde M_{h,33}^2 &=& -m_S^2 - \frac{1}{2}(m'^2_S -\kappa_1 \tilde \rho_1^2 -\kappa_2 \tilde \rho_2^2 -3\lambda_S \tilde s^2),\\
\tilde M_{h,12}^2 &=& -m_{12}^2 + \lambda_{345} \tilde \rho_1\tilde \rho_2,\quad
\tilde M_{h,13}^2 = \kappa_1 \tilde \rho_1 \tilde s,\quad
\tilde M_{h,23}^2 = \kappa_2 \tilde \rho_2 \tilde s.
\end{eqnarray}
The mass-squared matrix for the $CP$-odd neutral scalar bosons is
\begin{equation}
\tilde M_{0}^2=\begin{pmatrix}
m_{11}^2 + (\lambda_1\tilde \rho_1^2
+ \hat\lambda_{345} \tilde \rho_2^2 + \kappa_1 \tilde s^2)/2
&-m_{12}^2 + \lambda_5 \tilde \rho_1 \tilde \rho_2\\
-m_{12}^2 + \lambda_5 \tilde \rho_1 \tilde \rho_2
&m_{22}^2 + (\lambda_2 \tilde \rho_2^2
+ \hat\lambda_{345} \tilde \rho_1^2 + \kappa_2 \tilde s^2)/2
\end{pmatrix},
\end{equation}
with $\hat\lambda_{345} \equiv \lambda_3 + \lambda_4 - \lambda_5$,
while the mass-squared matrix for the charged scalar bosons is
\begin{equation}
\tilde M_{+}^2=\begin{pmatrix}
m_{11}^2 + (\lambda_1 \tilde \rho_1^2 + \lambda_3\tilde \rho_2^2 + \kappa_1 \tilde s^2)/2
&-m_{12}^2 + (\lambda_4 + \lambda_5) \tilde \rho_1 \tilde \rho_2/2\\
- m_{12}^2 + (\lambda_4 + \lambda_5) \tilde \rho_1 \tilde \rho_2/2
&m_{22}^2 + (\lambda_2 \tilde \rho_2^2 + \lambda_3 \tilde \rho_1^2 + \kappa_2 \tilde s^2)/2
\end{pmatrix}.
\end{equation}
The eigenvalues of these matrices give the masses squared of the scalar bosons, i.e.,
\begin{eqnarray}
\mathrm{eigenvalues}(\tilde M_h^2) &=& \left\{\tilde m_{h_1}^2,~\tilde m_{h_2}^2,~\tilde m_{h_3}^2\right\},\\
\mathrm{eigenvalues}(\tilde M_{0}^2) &=&  \left\{\tilde m_{G^0}^2,~\tilde m_{a}^2\right\},\\
\mathrm{eigenvalues}(\tilde M_{+}^2) &=& \left\{\tilde m_{G^\pm}^2,~\tilde m_{H^\pm}^2\right\}.
\end{eqnarray}
The masses squared of the DM candidate $\chi$ is obtained as
\begin{equation}
\tilde m_\chi^2 = -m_S^2 + \frac{1}{2} (m'^2_S + \kappa_1\tilde \rho_1^2 + \kappa_2\tilde \rho_2^2 + \lambda_S \tilde s^2).
\end{equation}
The mass squared of the $W^\pm$ boson is given by
\begin{equation}
\tilde m_{W^\pm_\mathrm{L}}^2 = \tilde m_{W^\pm_\mathrm{T}}^2 = \frac{g^2}{4}(\tilde \rho_1^2 + \tilde \rho_2^2).
\end{equation}
The mass-squared matrix of the $B$ and $W^3$ gauge fields is
\begin{equation}\label{eq:M2_W3B}
\tilde M^2_{W^3,B} = \frac{1}{4}(\tilde \rho_1^2 + \tilde \rho_2^2)
\begin{pmatrix}
g^2 & -gg'\\
-gg' & g'^2
\end{pmatrix}.
\end{equation}
After diagonalization, the masses squared of the $Z$ boson and the photon are
\begin{eqnarray}
\tilde m_{Z_\mathrm{L}}^2 &=& \tilde m_{Z_\mathrm{T}}^2 = \frac{1}{4}(g^2+g'^2)(\tilde \rho_1^2 + \tilde \rho_2^2),\\
\tilde m_{\gamma_\mathrm{T}}^2 &=& \tilde m_{\gamma_\mathrm{L}}^2=0.
\end{eqnarray}
For the type-I Yukawa couplings, the masses squared of the top and bottom quarks are
\begin{equation}
\tilde m_t^2= \frac{y_t^2}{2\sin^2\beta}\,\tilde \rho_2^2, \quad
\tilde m_b^2= \frac{y_b^2}{2\sin^2\beta}\,\tilde \rho_2^2,
\end{equation}
where the couplings $y_t = \sqrt{2}m_t/v$ and $y_b= \sqrt{2}m_b/v$ are defined the same as in the SM.
For the type-II Yukawa couplings, the masses squared become
\begin{equation}
\tilde m_t^2= \frac{y_t^2}{2\sin^2\beta}\,\tilde \rho_2^2, \quad
\tilde m_b^2= \frac{y_b^2}{2\cos^2\beta}\,\tilde \rho_1^2.
\end{equation}

Notice that loop corrections generally shift the values of the VEVs as well as the renormalized mass-squared matrix of the $CP$-even neutral scalar bosons. To keep them intact, we introduce the following counterterms~\cite{Cline:2011mm,Basler:2016obg},
\begin{align}\label{eq:CT}
V_{\mathrm{CT}}(\tilde \rho_1,\tilde \rho_2,\tilde s) &= \delta m_1^2\tilde \rho_1^2 + \delta m_2^2\tilde \rho_2^2 + \delta m_s^2{\tilde s^2} + \delta {\lambda_1}\tilde \rho_1^4 + \delta {\lambda_2}\tilde \rho_2^4 + \delta {\lambda_s}{\tilde s^4} \notag \\
& \quad + \delta \lambda_{12}\tilde \rho_1^2\tilde \rho_2^2 + \delta \lambda_{1s} \tilde \rho_1^2 \tilde s^2 + \delta \lambda_{2s}\tilde \rho_2^2 \tilde s^2.
\end{align}
The nine counterterm coefficients are determined by the following nine equations at $(\tilde \rho_1, \tilde \rho_2, \tilde s) = (v_1, v_2, v_s)$,
\begin{alignat}{3}
\frac{\partial V_{\mathrm{CT}}}{\partial\tilde \rho_1} &= - \frac{\partial V_1}{\partial \tilde \rho_1},\quad&
\frac{\partial V_{\mathrm{CT}}}{\partial \tilde \rho_2} &= - \frac{\partial V_1}{\partial \tilde \rho_2},\quad&
\frac{\partial V_1}{\partial \tilde s} &=  - \frac{\partial V_1}{\partial \tilde s},
\label{eq:2-orderCW:1}\\
\frac{\partial^2 V_{\mathrm{CT}}}{\partial \tilde \rho_1^2} &=  - \frac{\partial ^2 V_1}{\partial \tilde \rho_1^2},\quad&
\frac{\partial^2 V_{\mathrm{CT}}}{\partial \tilde \rho_2^2} &=  - \frac{\partial ^2 V_1}{\partial \tilde \rho_2^2},\quad&
\frac{\partial^2 V_{\mathrm{CT}}}{\partial \tilde s^2} &=  - \frac{\partial ^2 V_1}{\partial \tilde s^2},\quad
\label{eq:2-orderCW:2}\\
\frac{\partial^2 V_{\mathrm{CT}}}{\partial \tilde \rho_2\partial \tilde \rho_1} &=  - \frac{\partial^2 V_1}{\partial \tilde \rho_2\partial \tilde \rho_1},\quad&
\frac{\partial^2 V_{\mathrm{CT}}}{\partial \tilde s \partial \tilde \rho_1} &=  - \frac{\partial^2 V_1}{\partial \tilde s \partial \tilde \rho_1},\quad&
\frac{\partial^2 V_{\mathrm{CT}}}{\partial \tilde s \partial \tilde \rho_2} &=  - \frac{\partial^2 V_1}{\partial \tilde s \partial \tilde \rho_2}.
\label{eq:2-orderCW:3}
\end{alignat}

The masses of the Nambu-Goldstone bosons $G^0$ and $G^\pm$ vanish at $(\tilde \rho_1, \tilde \rho_2, \tilde s) = (v_1, v_2, v_s)$ in the Landau gauge, inducing logarithmic IR divergence terms in Eqs.~\eqref{eq:2-orderCW:2} and \eqref{eq:2-orderCW:3} proportional to
\begin{equation}
\frac{\partial \tilde m_G^2}{\partial \phi_i}\frac{\partial \tilde m_G^2}{\partial \phi_j}\ln \frac{\tilde m_G^2}{\mu^2},\quad
\phi_i = \tilde \rho_1,\tilde \rho_2,\tilde s.
\end{equation}
This problem is due to the ill-defined renormalized Higgs boson masses at $p^2=0$ with massless Nambu-Goldstone modes, and one can fix it by setting the momenta of the Higgs bosons on shell~\cite{Cline:1996mga,Casas:1994us}. Similar problems exist in the effective potential with higher loops, and more details can be found in Refs.~\cite{Elias-Miro:2014pca,Martin:2014bca}.
An approximate treatment is to give an IR cutoff $\Lambda_\mathrm{IR}$ to the Nambu-Goldstone boson masses~\cite{Cline:2011mm}, i.e., to set $\tilde m_{G^0}^2 = \tilde m_{G^\pm}^2 =  \Lambda_{\mathrm{IR}}^2$ in the logarithms at $(\tilde \rho_1, \tilde \rho_2, \tilde s) = (v_1, v_2, v_s)$.
Here, we take $\Lambda_\mathrm{IR}$ to be the mass of the SM-like Higgs boson.
Solving Eqs.~\eqref{eq:2-orderCW:1}--\eqref{eq:2-orderCW:3}, we obtain
\begin{eqnarray}
\delta m_1^2 &=&  - \frac{3}{4v_1}\frac{\partial V_1}{\partial \tilde \rho_1} + \frac{1}{4}\frac{\partial^2 V_1}{\partial \tilde \rho_1^2} + \frac{v_2}{4v_1}\frac{\partial^2 V_1}{\partial \tilde \rho_2 \partial \tilde \rho_1} + \frac{v_s}{4v_1}\frac{\partial^2 V_1}{\partial \tilde s \partial \tilde \rho_1},
\\
\delta m_2^2 &=&  - \frac{3}{4v_2}\frac{\partial V_1}{\partial \tilde \rho_2} + \frac{1}{4}\frac{\partial^2 V_1}{\partial \tilde \rho_2^2} + \frac{v_1}{4v_2}\frac{\partial^2 V_1}{\partial \tilde \rho_2 \partial \tilde \rho_1} + \frac{v_s}{4v_2}\frac{\partial^2 V_1}{\partial \tilde s \partial \tilde \rho_2},
\\
\delta m_s^2 &=&  - \frac{3}{4v_s}\frac{\partial V_1}{\partial \tilde s} + \frac{1}{4}\frac{\partial^2 V_1}{\partial \tilde s^2} + \frac{v_1}{4v_s}\frac{\partial^2 V_1}{\partial \tilde s \partial \tilde \rho_1} + \frac{v_2}{4v_s}\frac{\partial^2 V_1}{\partial \tilde s \partial \tilde \rho_2},
\\
\delta \lambda_1 &=& \frac{1}{8v_1^3}\frac{\partial V_1}{\partial \tilde \rho_1} - \frac{1}{8v_1^2}\frac{\partial^2 V_1}{\partial \tilde \rho_1^2},\quad
\delta \lambda_2 = \frac{1}{8v_2^3}\frac{\partial V_1}{\partial \tilde \rho_2} - \frac{1}{8v_2^2}\frac{\partial^2 V_1}{\partial \tilde \rho_2^2},
\\
\delta \lambda_s &=& \frac{1}{8v_s^3}\frac{\partial V_1}{\partial \tilde s} - \frac{1}{8v_s^2}\frac{\partial^2 V_1}{\partial \tilde s^2},\quad
\delta \lambda_{12} =  - \frac{1}{4v_1 v_2}\frac{\partial^2 V_1}{\partial \tilde \rho_2 \partial \tilde \rho_1},
\\
\delta \lambda_{1s} &=&  - \frac{1}{4v_1 v_s}\frac{\partial^2 V_1}{\partial \tilde s \partial \tilde \rho_1},\quad
\delta \lambda_{2s} =  - \frac{1}{4v_2 v_s}\frac{\partial^2 V_1}{\partial \tilde s \partial \tilde \rho_2},
\end{eqnarray}
at $(\tilde \rho_1, \tilde \rho_2, \tilde s) = (v_1, v_2, v_s)$.

Thermal corrections to the effective potential are crucial for studying the EWPT.
The one-loop finite-temperature effective potential~\cite{Dolan:1973qd} can be expressed as
\begin{equation}\label{eq:1loopT}
V_\mathrm{1T}(\tilde \rho_1,\tilde \rho_2,\tilde s,T) = \frac{T^4}
{2\pi^2}\left[ {\sum\limits_{i=\mathrm{bosons}} {n_i}{J_\mathrm{B}}\left( \frac{\tilde m_i^2}
{T^2} \right)}  + \sum\limits_{i={t,b}} {n_i}{J_\mathrm{F}}\left( \frac{\tilde m_i^2}{T^2} \right) \right],
\end{equation}
where $T$ is the temperature and the functions $J_\mathrm{B}$ and $J_\mathrm{F}$ are defined as
\begin{eqnarray}
J_\mathrm{B}(x) &\equiv& \int_0^\infty  {{y^2}\ln \left( {1 - e^{ - \sqrt {y^2 + x} }} \right)dy},
\\
J_\mathrm{F}(x) &\equiv& \int_0^\infty  {{y^2}\ln \left( {1 + e^{ - \sqrt {y^2 + x} }} \right)dy}.
\end{eqnarray}

We also consider the daisy diagrams, which can be significant.
The corressponding contribution to the effective potential can be estimated by~\cite{Carrington:1991hz,Arnold:1992rz}
\begin{equation}\label{eq:DaisyTerm}
V_{\mathrm{D}}(\tilde \rho_1,\tilde \rho_2,\tilde s,T) =  - \frac{T}{12\pi}\sum\limits_{i=\mathrm{bosons}} {n_i}\left[(\bar{m}_i^2)^{3/2} -(\tilde m_i^2)^{3/2} \right].
\end{equation}
$\bar{m}_i^2$ are the field-dependent boson masses squared with thermal corrections in the high-temperature limit and can be derived by
\begin{equation}
\bar{m}_i^2(\tilde \rho_1,\tilde \rho_2,\tilde s,T) = \mathrm{eigenvalues}[\tilde M_X^2(\tilde \rho_1,\tilde \rho_2,\tilde s)+\Pi_X(T)],
\end{equation}
where $\tilde M_X^2(\tilde \rho_1,\tilde \rho_2,\tilde s)$ represents the mass-squared matrices or masses squared in terms of the classical fields, and $\Pi_X (T)$ denotes the thermal corrections to $\tilde M_X^2$.
The subleading off-diagonal elements of $\Pi_X (T)$ can be neglected~\cite{Carrington:1991hz,Blinov:2015vma}.
The diagonal elements of $\Pi_X (T)$ for the scalar bosons are derived as
\begin{eqnarray}
\Pi_{h,11} &=& \Pi_{0,11} = \Pi_{+,11} = \frac{T^2}{48}(9{g^2} + 3{g'}^2 + 12\lambda_1 + 8\lambda_3 + 4\lambda_4 + 4\kappa_1 + y_1), \\
\Pi_{h,22} &=& \Pi_{0,22} = \Pi_{+,22} = \frac{T^2}{48}( 9{g^2} + 3{g'}^2 + 12\lambda_2 + 8\lambda_3 + 4\lambda_4 + 4\kappa_2 + y_2),  \\
\Pi_{h,33} &=& \Pi_\chi = \frac{T^2}{6}(\lambda_S + \kappa_1 + \kappa_2 ).
\end{eqnarray}
Here, $y_1$ and $y_2$ are the contributions from the Yukawa couplings.
For the type-I and -II cases, they are given by
\begin{eqnarray}
\text{Type I:}&&\quad
y_1 = 0,\quad
y_2 = \frac{12(y_t^2 + y_b^2)}{\sin^2\beta},
\\
\text{Type II:}&&\quad
y_1 = \frac{12 y_b^2}{\cos^2\beta},\quad
y_2 = \frac{12 y_t^2}{\sin^2\beta}.
\end{eqnarray}
The thermal corrections to the electroweak gauge bosons are
\begin{eqnarray}
\Pi_{W^\pm_\mathrm{L}} &=& \Pi_{W^3_\mathrm{L}} = 2g^2 T^2, \\ \Pi_{B_\mathrm{L}} &=& 2{g'}^2 T^2,\\
\Pi_{W^\pm_\mathrm{T}} &=& \Pi_{Z_\mathrm{T}} = \Pi_{\gamma_\mathrm{T}} = 0.
\end{eqnarray}
Note that $\Pi_{W^3_\mathrm{L}}$ and $\Pi_{B_\mathrm{L}}$ are the corrections to the diagonal elements of $\tilde M^2_{W^3,B}$ in Eq.~\eqref{eq:M2_W3B}.

Finally, we obtain the total effective potential\footnote{Discussions on theoretical uncertainties in perturbative calculations of the effective potential can be found in Ref.~\cite{Croon:2020cgk}.}
\begin{equation}
V_\mathrm{eff}(\tilde \rho_1,\tilde \rho_2,\tilde s,T) = V_0 + V_1 + V_{\mathrm{CT}} + V_\mathrm{1T} + V_{\mathrm{D}}.
\end{equation}

\section{Phase transitions}
\label{sec:pts}

Based on the effective potential constructed in the previous section, we can study its evolution with temperature.
At sufficiently high temperatures, the effective potential is minimized at the origin $(\tilde\rho_1, \tilde\rho_2, \tilde s) = (0,0,0)$, implying the restoration of the electroweak gauge symmetry.
As the Universe cools down, extra minima appear.
In particular, if there are two coexisted minima separated by a high barrier, strong FOPT could take place and result in a stochastic GW background.
We utilize the numerical package \texttt{CosmoTransitions}~\cite{Wainwright:2011kj} to analyze the phase transitions.
For each parameter point in the random scans, we verify whether or not the minimum $(\tilde \rho_1, \tilde \rho_2, \tilde s) = (v_1, v_2, v_s)$ is the global one of the zero-temperature effective potential.
The parameter points that fail this test are rejected.
Then we use \texttt{CosmoTransitions} to trace the temperature evolution of the local minima.

\begin{figure}[!t]
	\centering
	\subfigure[~Minima in the $T$-$v_1(T)$ plane.\label{fig:phase_rho1}]
	{\includegraphics[width=0.48\textwidth]{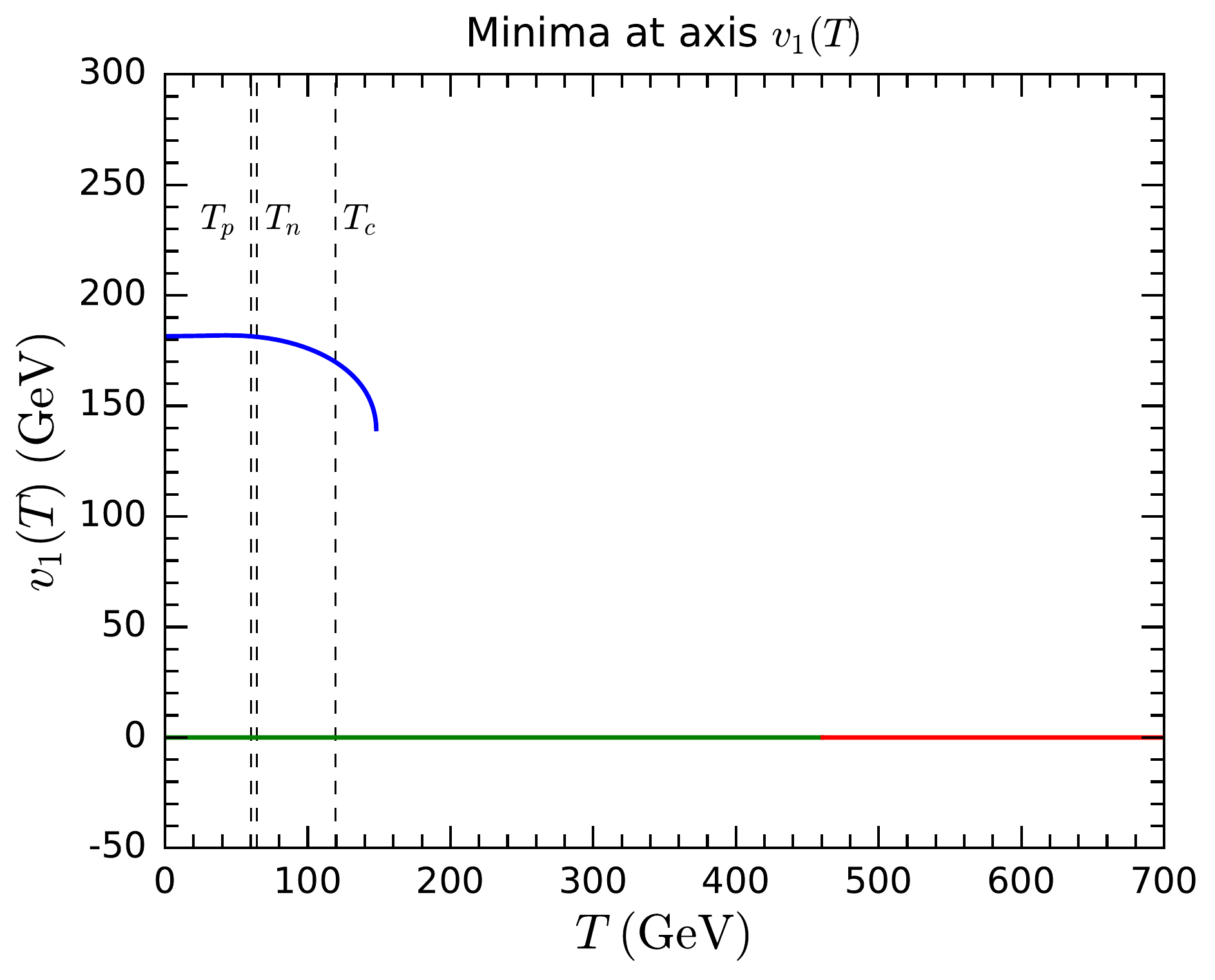}}
	\subfigure[~Minima in the $T$-$v_2(T)$ plane.\label{fig:phase_rho2}]
	{\includegraphics[width=0.48\textwidth]{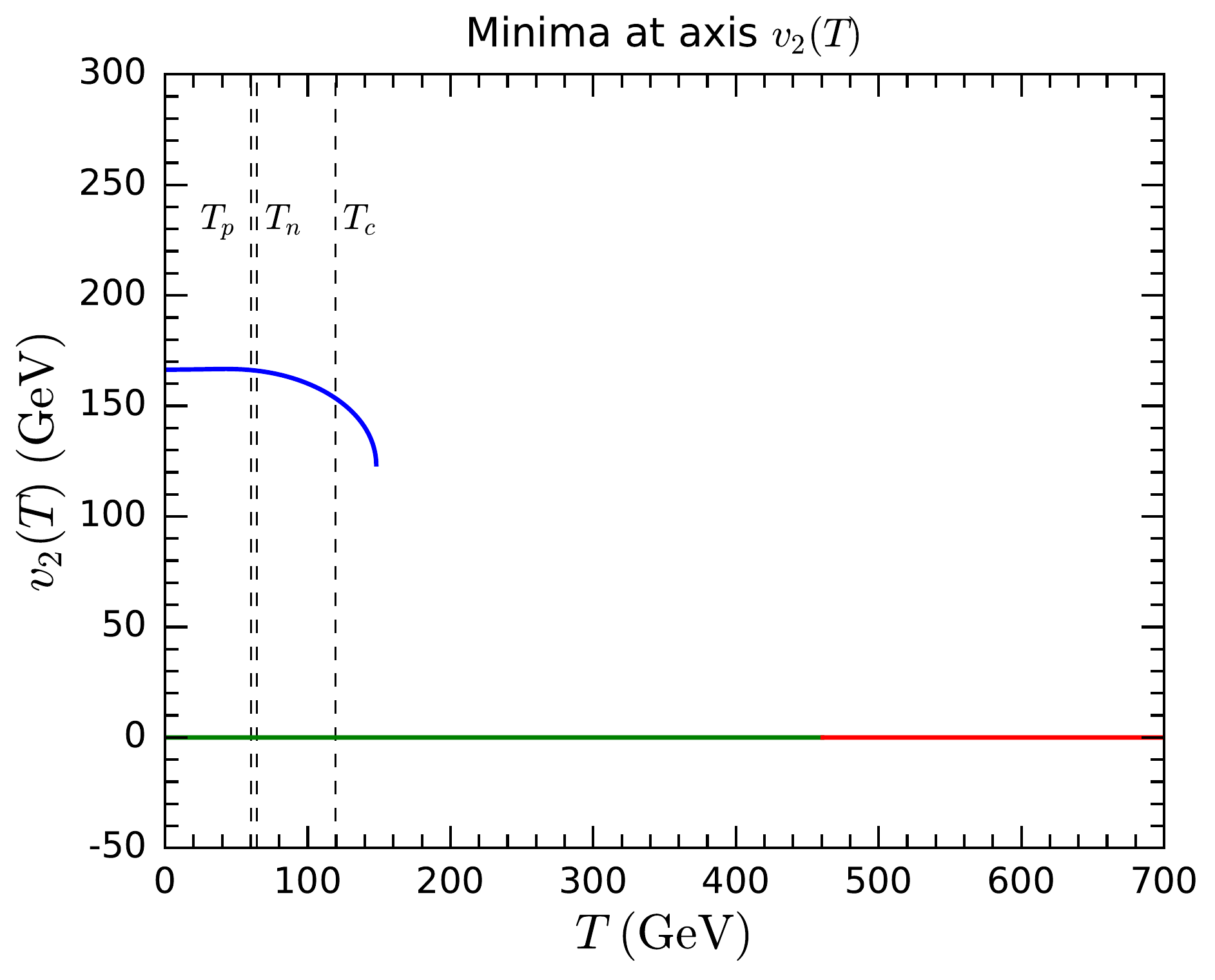}}
	\subfigure[~Minima in the $T$-$v_s(T)$ plane.\label{fig:phase_s}]
	{\includegraphics[width=0.48\textwidth]{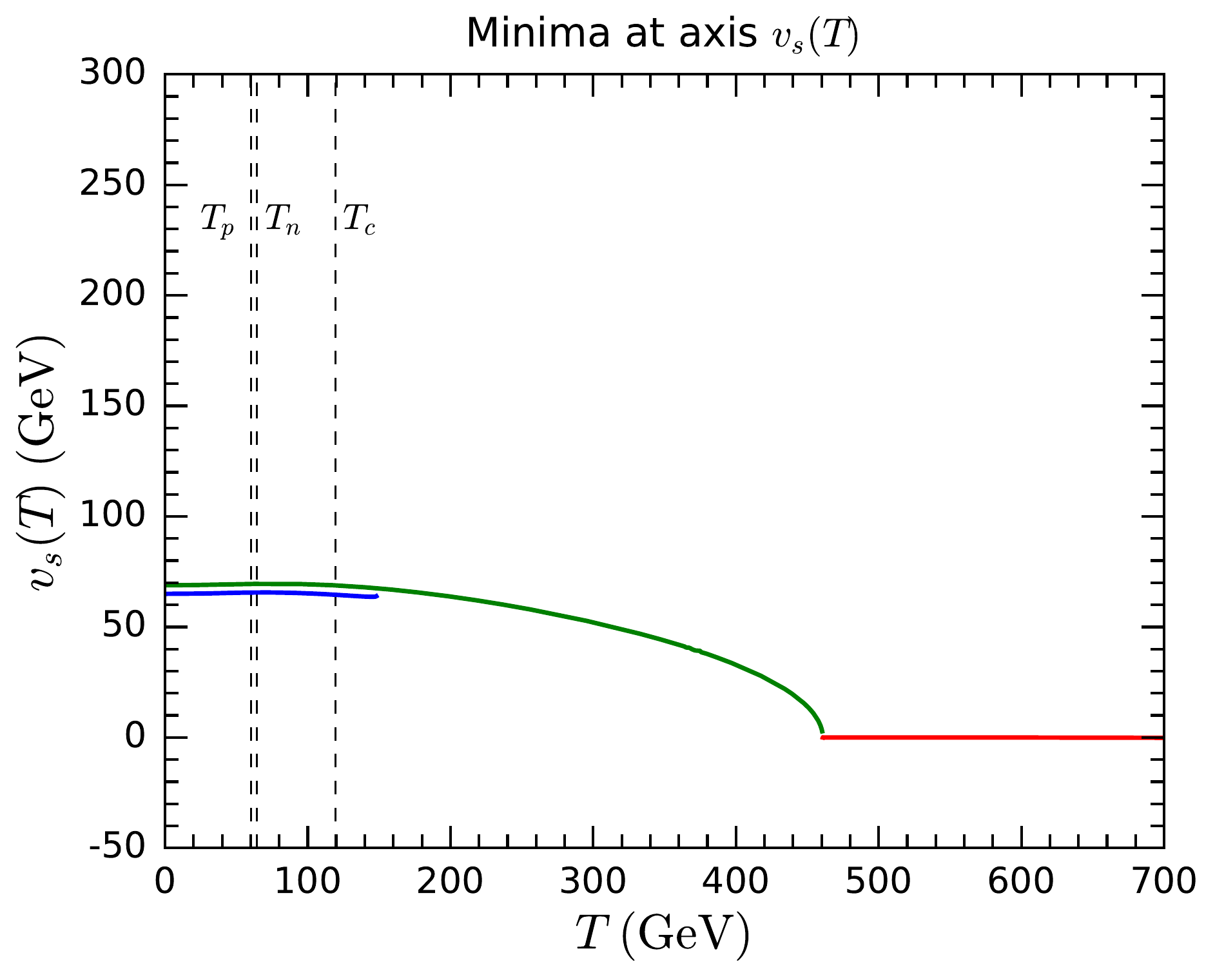}}
	\caption{Temperature evolution of the positions of the minima in the axes $v_1(T)$, $v_2(T)$, and $v_s(T)$ for BP3.
    The red, green, and blue lines denote three local minima.
    The vertical dashed lines indicate the critical, nucleation, and percolation temperatures $T_\mathrm{c}$, $T_\mathrm{n}$, and $T_\mathrm{p}$.}
	\label{fig:phases}
\end{figure}

In this model, the three classical $CP$-even neutral scalar fields would develop VEVs, typically leading to multi-step cosmological phase transitions.
In Fig.~\ref{fig:phases}, we demonstrate the temperature evolution of multiple phases for a benchmark point (BP), whose parameters can be found in the BP3 column of Table~\ref{tab:benchmarks} in Sec.~\ref{sec:num}.
In the plots, $v_1(T)$, $v_2(T)$, and $v_s(T)$ are the $T$-dependent values of the classical fields $\tilde\rho_1$, $\tilde\rho_2$, and $\tilde s$ at the local minima of the effective potential.
The red, green, and blue lines indicate the positions of three local minima.

At $T \gtrsim 460~\si{GeV}$, the system stays at the red minimum with $\big(v_1(T), v_2(T), v_s(T)\big) = (0,0,0)$, respecting the electroweak gauge symmetry.
At $T \simeq 460~\si{GeV}$, a second-order phase transition occurs and the system turns into the green minimum, where $\tilde s$ develops a nonzero VEV.
At $T \simeq 148~\si{GeV}$, the blue minimum appears, accompanied with a barrier that separates it from the green minimum.
These two minima coexist till the zero temperature.

The effective potential at the blue minimum is higher than at the green minimum until the critical temperature $T_\mathrm{c} \simeq 119~\si{GeV}$.
Below $T_\mathrm{c}$, the green minimum becomes a metastable state, i.e., a ``false vacuum''.
The system finally undergoes a FOPT through quantum tunneling and turns into the blue minimum,  or the ``true vacuum''.
Such a FOPT nucleates bubbles, inside which the system is trapped at the true vacuum.
In this FOPT, $v_1(T)$ and $v_2(T)$ increase from zero to $\mathcal{O}(100)~\si{GeV}$, while $v_s(T)$ slightly decreases.
At zero temperature, the true vacuum satisfies $\big(v_1(0), v_2(0), v_s(0)\big) = (v_1, v_2, v_s)$.

In our parameter scans, we usually find that $v_s(T)$ does not evolve synchronously with $v_1(T)$ and $v_2(T)$, probably due to the less couplings of the singlet field to other fields compared with the two Higgs doublets.
Typically, $v_s(T)$ becomes nonzero much earlier than the conventional EWPT epoch via a second-order or first-order phase transition. $\tilde\rho_1$ and $\tilde\rho_2$ then gain VEVs in an subsequent phase transition, which could be a strong FOPT similar to those in the conventional two-Higgs-doublet models~\cite{Dorsch:2017nza,Bernon:2017jgv}.

Below we discuss the dynamics of the FOPTs.
The bubble nucleation rate per unit time and unit volume is given by~\cite{Linde:1980tt,Linde:1981zj}
\begin{equation}
\Gamma \sim AT^4e^{-S},
\end{equation}
where $A$ is an $\mathcal{O}(1)$ constant and $S = \mathrm{min}\{S_4,S_3/T\}$.
$S_4$ and $S_3$ are the Euclidean actions of the scalar fields for $O(4)$- and $O(3)$-symmetric bubbles, respectively.  The three-dimensional action $S_3$ can be simplified to
\begin{equation}
S_3 = 4\pi \int_0^\infty  dr\, r^2 \left[ {\frac{1}{2}{\frac{d \phi_i }{dr}}{\frac{d \phi_i }{dr}} + V_\mathrm{eff}(\phi_i ,T)} \right],
\end{equation}
where $r$ is the radius of the bubble.
$\phi_i(r) =
\big(\tilde\rho_1(r),\tilde\rho_2(r),\tilde s(r)\big)$ is given by the bounce solution of the equations of motion
\begin{equation}
\frac{d^2 \phi_i }{dr^2} + \frac{2}{r}\frac{d \phi_i}{dr} = \frac{\partial V_\mathrm{eff}}{\partial \phi_i}
\end{equation}
with boundary conditions
\begin{equation}
\left. \frac{d\phi_i }{dr} \right|_{r = 0} = 0,\quad \phi_i (\infty ) = \phi_i^{\mathrm{false}},
\end{equation}
where $\phi_i^{\mathrm{false}}$ is the field configuration of the false vacuum.

\begin{figure}[!t]
	\centering
    \subfigure[~$S_3/T$ and $S_4$.\label{fig:action:S}]
	{\includegraphics[width=0.48\textwidth]{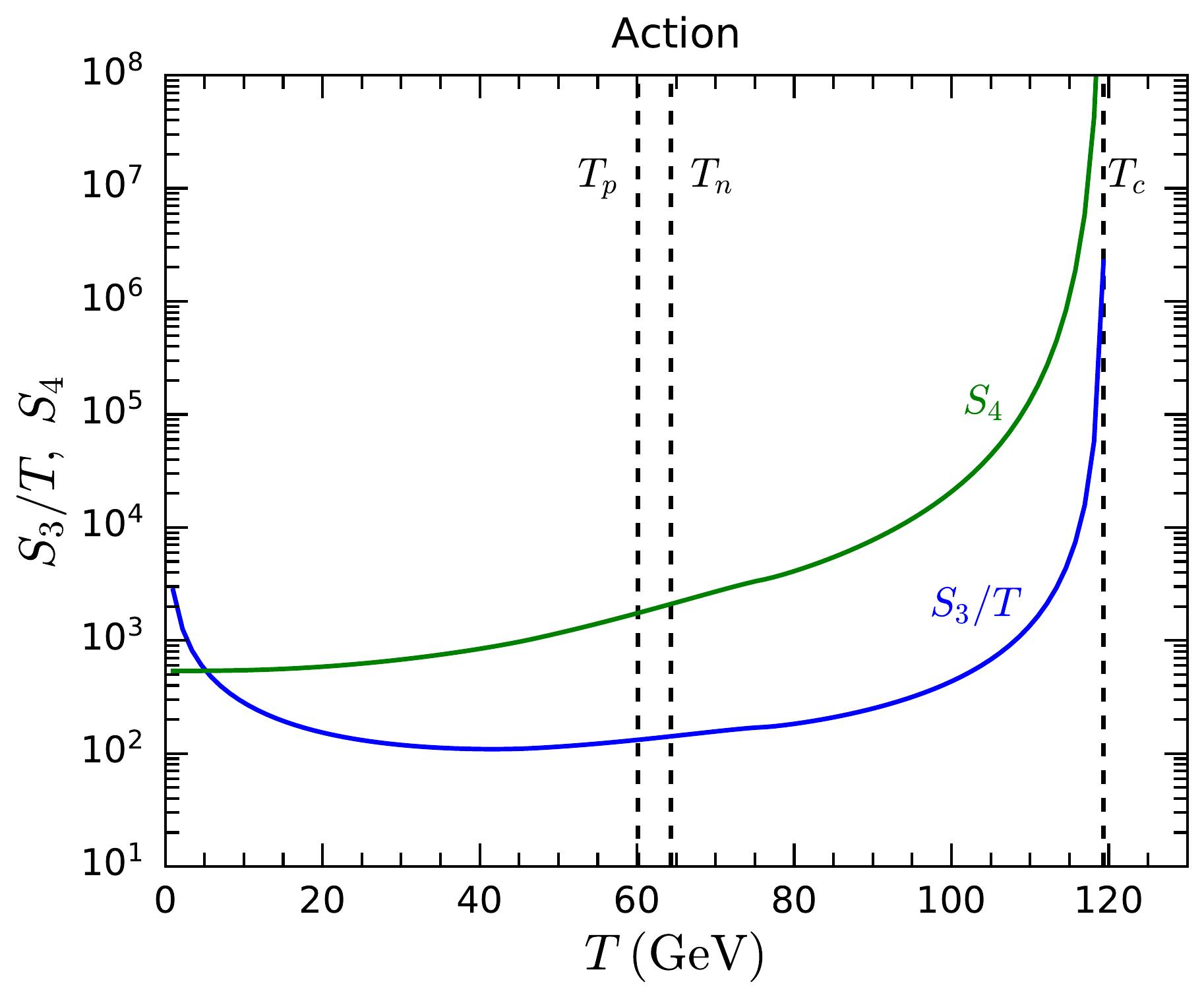}}
    \subfigure[~Nucleation rate $\Gamma$.\label{fig:action:Gamma}]
	{\includegraphics[width=0.48\textwidth]{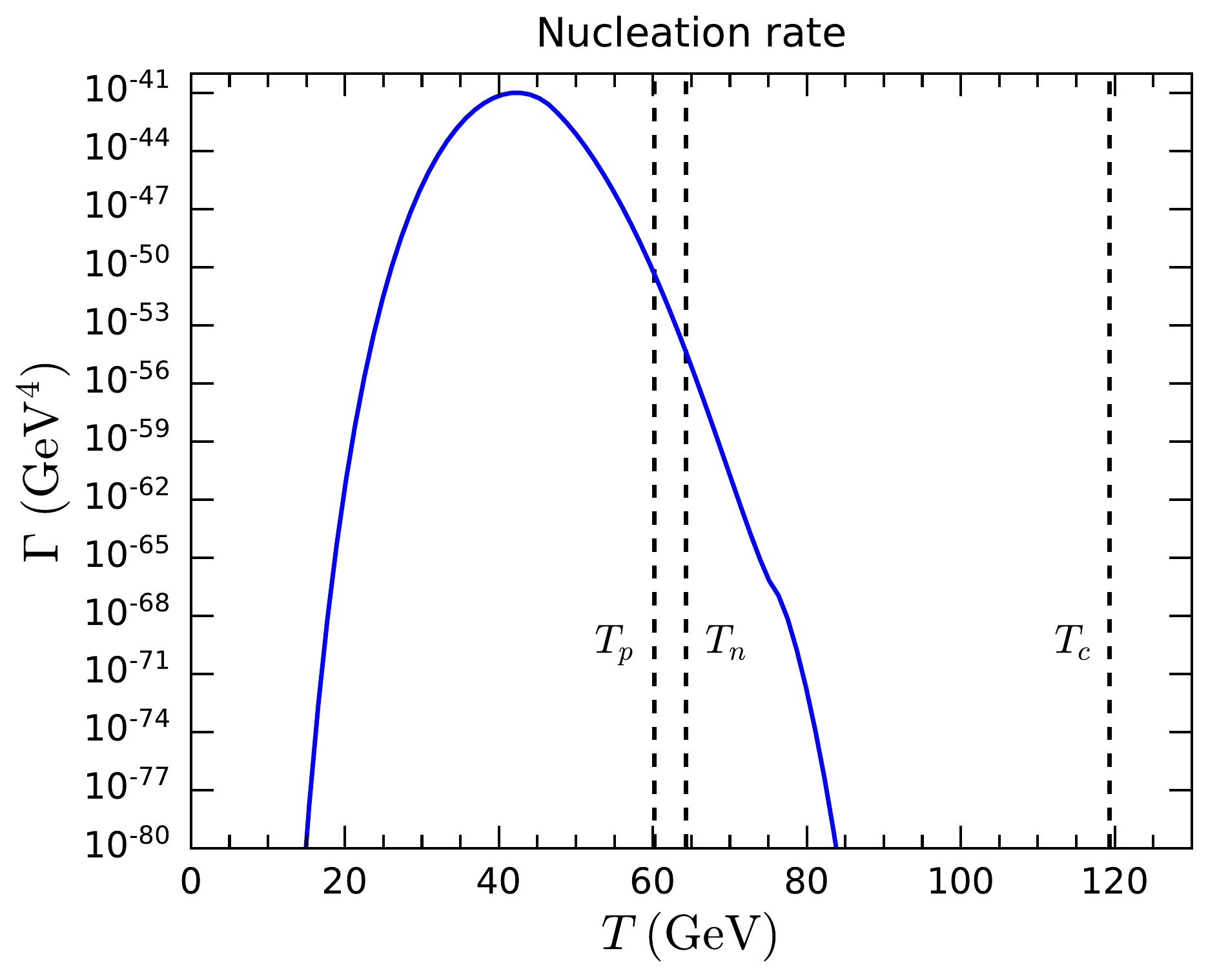}}
	\caption{The actions $S_3/T$ and $S_4$ (a) and the nucleation rate $\Gamma$ (b) as functions of the temperature $T$ for BP3.
    The dashed lines denote the critical, nucleation, and percolation temperatures $T_\mathrm{c}$, $T_\mathrm{n}$, and $T_\mathrm{p}$.}
	\label{fig:action}
\end{figure}

We present $S_4$ and $S_3/T$ as functions of the temperature for BP3 in Fig.~\ref{fig:action:S}, as well as the corresponding nucleation rate $\Gamma$ in Fig.~\ref{fig:action:Gamma}.
We find that $S_3/T$ is the smaller one until temperatures below $\sim 5~\si{GeV}$.
The minimal of $S_3/T$ is reached at $T \sim 40~\si{GeV}$.
As the Universe cools down, below the critical temperature $T_\mathrm{c}$, the nucleation rate increases before the peak around $T \sim 40~\si{GeV}$, and then decreases.

The bubbles are actually nucleated at the nucleation temperature $T_\mathrm{n}$, where the nucleation probability for a single bubble within a Hubble volume reaches $\mathcal{O} (1)$.
Thus, $T_\mathrm{n}$ can be estimated by~\cite{Moreno:1998bq}
\begin{equation}\label{eq:Tn}
\int_{t_\mathrm{c}}^{t_\mathrm{n}}{dt\,\frac{\Gamma }{H^3}} = \int_{T_\mathrm{n}}^{T_\mathrm{c}}{dT\,\frac{\Gamma }{H^4 T}} = 1,
\end{equation}
where $H$ is the Hubble rate, and $t_\mathrm{c}$ and $t_\mathrm{n}$ denote the critical and nucleation times, respectively.
Note that the differential relation between the time $t$ and the temperature $T$ is $dt = -(HT)^{-1} dT$ in the radiation-dominated epoch.

Below the nucleation temperature $T_\mathrm{n}$, an increasing number of bubbles thrive and collide with each other.
The maximum of bubble collisions that remarkably produces stochastic GWs is expected to be reached when percolation occurs~\cite{Leitao:2012tx}.
In order to evaluate the percolation time $t_\mathrm{p}$, we need to estimate the fraction of space that still remains in the false vacuum at time $t$, which can be computed by~\cite{Guth:1979bh,Guth:1981uk}
\begin{equation}\label{eq:Tp}
P(t) = \exp\left[-\frac{4\pi}{3}\int_{t_\mathrm{c}}^{t}{dt'\, \Gamma(t')\, a^3(t')\,r^3(t, t')}\right],
\end{equation}
where $a(t')$ is the scale factor. $r(t, t')$ is the comoving radius of a bubble growing from $t'$ to $t$, given by
\begin{equation}
r(t,t') = \int_{t'}^{t} {d\tau} \,\frac{v_\mathrm{w}}{a(\tau)},
\end{equation}
where $v_\mathrm{w}$ is the velocity of the bubble wall.
For randomly distributed spherical bubbles with equal size in the three-dimensional space, the percolation threshold is reached when the fraction of space converted to the true vacuum, $1 - P(t)$, increases to $\sim 0.29$~\cite{Shante_1971,Rintoul_1997}.
Thus, the percolation time $t_\mathrm{p}$ can be derived by requiring $P(t_\mathrm{p}) \simeq 0.71$, with the corresponding temperature $T_\mathrm{p}$ characterizing GW production from FOPTs~\cite{Leitao:2012tx,Kobakhidze:2017mru,Ellis:2018mja,Wang:2020jrd}.

FOPTs are able to release latent heat from the vacuum energy, which drives the expansion of the bubbles and also converts into the thermal and bulk kinetic energies of the plasma~\cite{Steinhardt:1981ct,Kamionkowski:1993fg,Espinosa:2010hh}.
The density of the released vacuum energy is given by~\cite{Enqvist:1991xw}
\begin{equation}
\rho_\mathrm{vac} = V_\mathrm{eff}(\phi_i^{\mathrm{false}},T) - V_\mathrm{eff}(\phi_i^{\mathrm{true}},T) - T\frac{\partial}{\partial T}[ V_\mathrm{eff}(\phi_i^{\mathrm{false}},T) - V_\mathrm{eff}(\phi_i^{\mathrm{true}},T)],
\end{equation}
where $\phi_i^{\mathrm{true}}$ is the field configuration of the true vacuum.
It is useful to define a dimensionless strength parameter
\begin{equation}
\alpha \equiv \frac{\rho_\mathrm{vac}}{\rho_{\mathrm{rad}}},
\end{equation}
with $\rho_{\mathrm{rad}} = {{\pi^2}{g_*}T^4}/{30}$ the radiation energy density in the plasma.
$g_*$ is the effective relativistic degrees of freedom in the plasma.

The expansion of the bubbles depends on the interactions between the bubble walls and the plasma, analogous to chemical combustion in a relativistic fluid~\cite{Steinhardt:1981ct}.
Hydrodynamic analyses show that bubble propagation have diverse modes, including Jouguet detonations, weak detonations, subsonic deflagrations, supersonic deflagrations (hybrid), and runway bubble walls~\cite{Espinosa:2010hh}.
Thus, it is difficult to completely work out the bubble wall velocity $v_\mathrm{w}$.
For Jouguet detonations, the Chapman-Jouguet condition leads to a wall velocity of~\cite{Steinhardt:1981ct}
\begin{equation}
v_\mathrm{CJ} = \frac{1  + \sqrt {3\alpha ^2 + 2\alpha } }{\sqrt{3}(1 + \alpha) }.
\end{equation}
This is a typical assumption when evaluating GW signals.

Expanding the action $S$ around the time $t' = t_\mathrm{n}$ or $t' = t_\mathrm{p}$, we have
\begin{equation}
S(t)\simeq S(t')- \beta (t-t') + \mathcal{O}[(t-t')^2],
\end{equation}
where
\begin{equation}
\beta \equiv \left. -\frac{dS}{dt} \right|_{t=t'} = \left. \left(HT \frac{dS}{dT}\right) \right|_{T=T'}
\end{equation}
can be roughly understood as the inverse time duration of the phase transition~\cite{Kosowsky:1991ua}.
For the electroweak FOPTs in which we are interested, the derivative $dS/dT$ is positive at $T = T'$, leading to positive $\beta$.
In addition, $S_3/T$ is typically smaller than $S_4$ at $T = T'$.
In order to conveniently compare the phase transition time scale $\beta^{-1}$ and the cosmological expansion time scale $H^{-1}$, we define a dimensionless quantity
\begin{equation}
\tilde{\beta}(T')\equiv \frac{\beta(T')}{H(T')}.
\end{equation}

Based on Eqs.~\eqref{eq:Tn} and \eqref{eq:Tp}, further calculations show that the nucleation and percolation temperatures $T_\mathrm{n}$ and $T_\mathrm{p}$ can be approximately determined by~\cite{Huber:2007vva}
\begin{eqnarray}
\frac{S_3(T_\mathrm{n})}{T_\mathrm{n}} &\simeq& 141.5 - 2\ln\frac{g_*}{100} - 4\ln \frac{T_\mathrm{n}}{100~\si{GeV}} - \ln \frac{\tilde\beta(T_\mathrm{n})}{100},\\
\frac{S_3(T_\mathrm{p})}{T_\mathrm{p}} &\simeq& 132.0 - 2\ln\frac{g_*}{100} - 4\ln\frac{T_\mathrm{p}}{100~\si{GeV}}  - 4\ln\frac{\tilde\beta(T_\mathrm{p})}{100} + 3\ln v_\mathrm{w}.
\end{eqnarray}
For BP3, the nucleation temperature is $T_\mathrm{n} \simeq 64~\si{GeV}$, while the percolation temperature assuming $v_\mathrm{w} = v_\mathrm{CJ}$ is slightly lower, $T_\mathrm{p} \simeq 60~\si{GeV}$, as denoted in Figs.~\ref{fig:phases} and \ref{fig:action}.

\section{Gravitational wave spectra}
\label{sec:gws}

Electroweak FOPTs could induce significant perturbations of the Friedmann-Robertson-Walker metric and produce stochastic GWs around the mHz band.
Two key parameters relevant to the relic GW spectrum are $\alpha$ and $\tilde{\beta}$ evaluated at the time $t_*$ when GWs are produced.
There are three coexisting GW sources at a FOPT, namely bubble collisions, sound waves, and magnetohydrodynamic (MHD) turbulence~\cite{Binetruy:2012ze,Caprini:2015zlo,Caprini:2019egz,Hindmarsh:2020hop}.
Denoting $\Omega_{\mathrm{GW}}$ to be
the present GW energy density per logarithmic frequency interval divided by the critical density,
we separate the contributions from the three sources as
\begin{equation}
\Omega_{\mathrm{GW}}h^2 = \Omega_{\mathrm{col}}h^2 + \Omega_{\mathrm{sw}}h^2 + \Omega_{\mathrm{turb}}h^2.
\end{equation}

\textbf{(a) Bubble collisions}

The nucleated bubbles expand and finally collide with each other.
Their collisions break the spherical symmetry and generate gravitational waves~\cite{Kosowsky:1991ua}.
This process can be well described by the envelope approximation~\cite{Kosowsky:1991ua,Kosowsky:1992rz,Kosowsky:1992vn}.
Numerical simulations for bubble collisions in the thermal plasma~\cite{Kamionkowski:1993fg,Huber:2008hg} show that the resulting GW spectrum at present can be approximated by
\begin{equation}\label{eq:Omega_col}
\Omega_{\mathrm{col}}h^2 = 1.67\times 10^{-5} \
\frac{0.11 v_\mathrm{w}^3}{(0.42+v_\mathrm{w}^2)\tilde{\beta}^2}  \left( \frac{\kappa_{\phi} \alpha}{1 + \alpha} \right)^2 \left( \frac{100}{g_*} \right)^{1/3} \frac{3.8(f/f_{\mathrm{col}})^{2.8}}{1 + 2.8(f/f_{\mathrm{col}})^{3.8}},
\end{equation}
where $g_*$ is evaluated at $T=T_*$, the temperature corresponding to $t=t_*$.
The peak frequency of the spectrum can be modeled as~\cite{Huber:2008hg}
\begin{equation}
f_{\mathrm{col}} = \frac{0.62\,\tilde{\beta}h_*}{1.8 - 0.1 v_\mathrm{w} + v_\mathrm{w}^2}.
\end{equation}
The redshift of the frequency has been taken into account by the factor
\begin{equation}
h_* = \frac{a(t_*)H(t_*)}{a(t_0)} = 1.65 \times 10^{-5}\ \mathrm{Hz}\  \frac{T_*}{100\ \mathrm{GeV}}\left( \frac{g_*}{100} \right)^{1/6},
\end{equation}
which is the inverse Hubble time at $t=t_*$ redshifted to today ($t=t_0$).
The efficiency factor $\kappa_{\phi}$ characterizes the fraction of the available vacuum energy converted into the gradient energy of the scalar fields.

\textbf{(b) Sound waves}

The explosive bubble expansion in the plasma induces a sound shell around the bubble wall.
After the bubble collisions, the sound shells propagate into the fluid as sound waves, which become a significant GW source~\cite{Hindmarsh:2013xza,Hindmarsh:2015qta,Hindmarsh:2017gnf}.
This source lasts until the sound waves are disrupted by the development of nonlinear shocks and turbulence~\cite{Hindmarsh:2017gnf,Ellis:2018mja,Ellis:2019oqb,Ellis:2020awk}.
Therefore, the duration of the sound wave source can be determined by the nonlinearity timescale estimated as~\cite{Ellis:2020awk}
\begin{equation}
\tau_{\mathrm{nl}}\sim \frac{(8\pi)^{1/3}v_\mathrm{w}}{\tilde{\beta} H_*}\sqrt{\frac{4(1+\alpha)}{3\kappa_v \alpha}},
\end{equation}
where $H_* \equiv H(t_*)$ is the Hubble rate at $t = t_*$ and $\kappa_v$ is the fraction of the available vacuum energy converted into the kinetic energy of the fluid bulk motion.
For Jouguet detonations, $v_\mathrm{w} = v_\mathrm{CJ}$, and $\kappa_v$ can be approximated by~\cite{Espinosa:2010hh}
\begin{equation}
\kappa_v^\mathrm{CJ} = \frac{\sqrt{\alpha}}{0.135 + \sqrt{ 0.98 + \alpha}}.
\end{equation}

In an expanding radiation-dominated Universe, the finite duration of the sound wave source leads to a suppression factor~\cite{Guo:2020grp}
\begin{equation}
\Upsilon =1-\frac{1}{\sqrt{1 + 2 \tau_\mathrm{nl} H_{*} }}.
\end{equation}
Thus, the GW spectrum contributed by the sound waves is given by~\cite{Hindmarsh:2017gnf,Guo:2020grp}
\begin{equation}\label{eq:Omega_sw}
\Omega_{\mathrm{sw}}h^2 = 1.17\times 10^{-6}\  \frac{\Upsilon v_\mathrm{w}}{\tilde{\beta}} \left( \frac{\kappa_v \alpha}{1 + \alpha} \right)^2 \left( \frac{100}{g_*} \right)^{1/3} \left( \frac{f}{f_{\mathrm{sw}}} \right)^3 \left( \frac{7}{4 + 3f^2/f_{\mathrm{sw}}^2} \right)^{7/2},
\end{equation}
where the peak frequency is estimated to be~\cite{Hindmarsh:2017gnf}
\begin{equation}
f_{\mathrm{sw}} = \frac{0.54\tilde{\beta}h_*}{v_\mathrm{w}}.
\end{equation}

\textbf{(c) MHD turbulence}

Bubble collisions can stir up turbulence in the fluid, as the energy injection to the plasma results in an extremely high Reynolds number~\cite{Kamionkowski:1993fg}.
Since the plasma is fully ionized, the magnetic field, along with the velocity field, should be considered, leading to MHD turbulence~\cite{Kosowsky:2001xp}.
It takes several Hubble times for the MHD turbulence to decay, and the stochastic GWs arise continuously during this period~\cite{Caprini:2009yp}.
The corresponding GW spectrum can be fitted as~\cite{Caprini:2009yp,Caprini:2015zlo}
\begin{equation}\label{eq:Omega_turb}
\Omega_{\mathrm{turb}}h^2 = 3.35 \times 10^{-4}\  \frac{v_\mathrm{w}}{\tilde{\beta}} \left(\frac{\kappa_{\mathrm{turb}}\alpha }{1 + \alpha } \right)^{3/2}  \left(\frac{100}{g_*} \right)^{1/3} \frac{(f/f_{\mathrm{turb}})^3}{(1 + f/f_{\mathrm{turb}})^{11/3}(1 + 8\pi f/{h_*})},
\end{equation}
with
\begin{equation}
f_{\mathrm{turb}} = \frac{3.5\tilde{\beta}h_*}{2v_\mathrm{w}}.
\end{equation}
Based on the suggestion from simulations, we optimistically set $\kappa_{\mathrm{turb}} \simeq 0.1\kappa_v$~\cite{Hindmarsh:2015qta,Caprini:2015zlo}.

In general, the contribution from the sound waves dominates in the GW spectrum~\cite{Hindmarsh:2020hop}.
Moreover, $\kappa_{\phi}$ is typically negligible, except for runaway bubble walls~\cite{Espinosa:2010hh,Caprini:2015zlo,Ellis:2019oqb}.
Thus, we omit the contribution from the bubble collisions in the following calculations.

\section{Numerical analyses}
\label{sec:num}

We perform random scans with the model parameters in the ranges of $v_s \in [10,~1000]~\si{GeV}$, $m_{\chi} \in [58,~800]~\si{GeV}$, $|m^2_{12}| \in [1,~500^2]~\si{GeV^2}$, $\tan\beta \in [0.5,~20]$, $\lambda_1$, $\lambda_2$, $|\lambda_3|$, $|\lambda_4|$, $|\lambda_5|$, $\lambda_S \in [0.8,~8]$, and $|\kappa_1|$, $|\kappa_2| \in [0.01,~8]$ for the two types of Yukawa couplings.
We assume that the prior probabilities for the random parameters follow uniform distributions in the logarithmic scale.
The parameter points are required to pass all the experimental constraints described in Sec.~\ref{sec:bound}, as well as to cause a FOPT.

Note that positive $\lambda_1$, $\lambda_2$, and $\lambda_S$ are required to satisfy the bounded-from-below conditions \eqref{eq:positive_lambda_12S}.
A negative $v_s$ would be totally equivalent to a positive one due to the $Z_2$ symmetry $S \to -S$.
Besides, a parameter point with $\tan\beta$ and $m_{12}^2$ is equivalent to one with $- \tan\beta$ and $- m_{12}^2$, since the potential respects the $Z_2$ symmetry $\Phi_1 \to -\Phi_1$ or $\Phi_2 \to -\Phi_2$ expect for the soft breaking quadratic terms with $m_{12}^2$.
Thus, we can just take positive $v_s$ and $\tan\beta$ in the scans, while $m_{12}^2$, $\lambda_3$, $\lambda_4$, $\lambda_5$, $\kappa_1$, and $\kappa_2$ can be either positive or negative.
In addition, the $v_s$ range of  $10~\si{GeV}$ to $1~\si{TeV}$ ensures that $S$ has a VEV near the electroweak scale, and the interplay between $S$ and the two Higgs doublets could be important.

\begin{figure}[!t]
	\centering	
	{\includegraphics[width=0.5\textwidth]{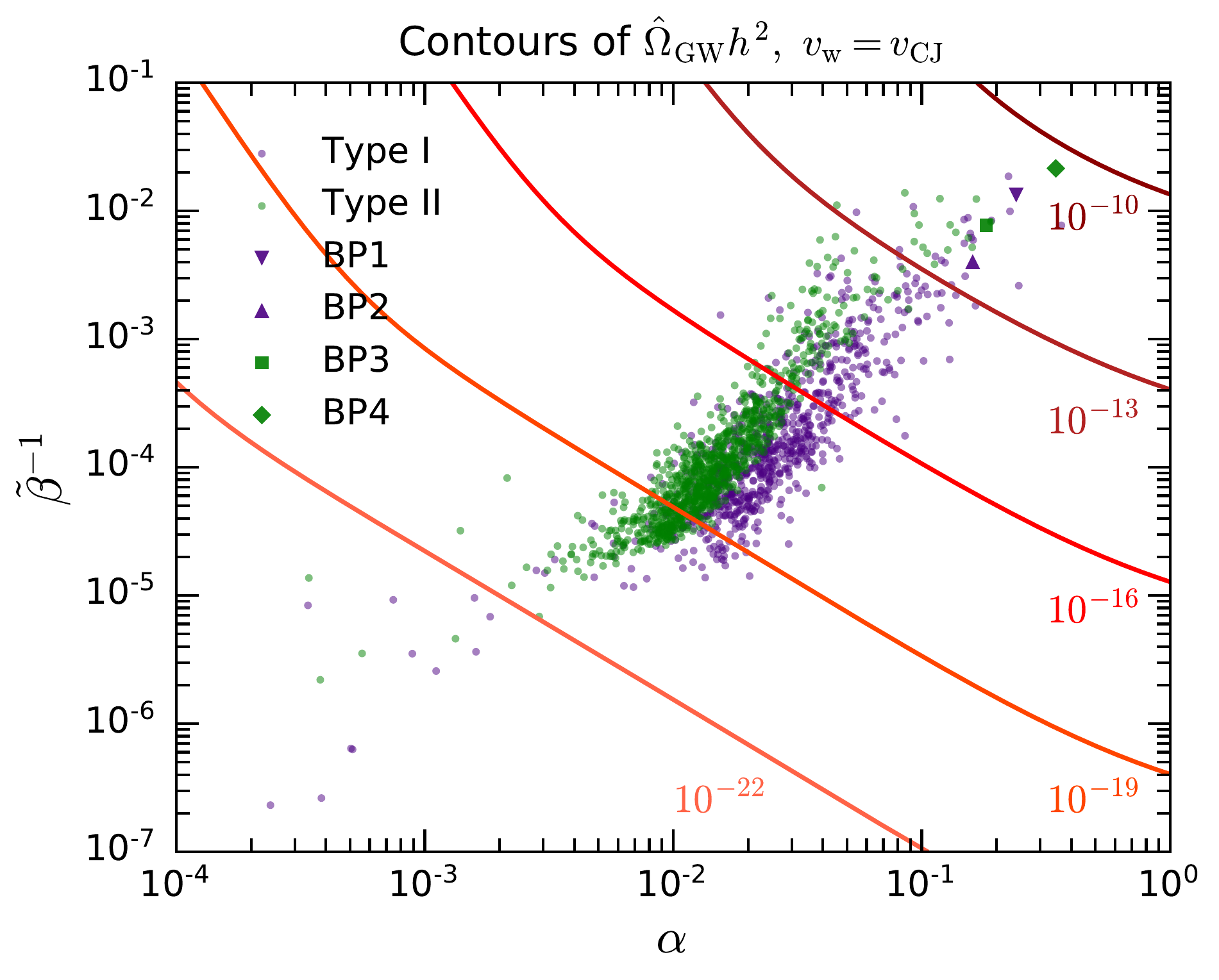}}
\caption{Contours of the peak amplitudes of the GW spectra in the $\alpha$-$\tilde\beta^{-1}$ plane assuming Jouguet detonations.
Purple and green points denote the parameter points for type-I and type-II Yukawa couplings, respectively.
Four BPs are also indicated.}
\label{fig:alpha_beta}
\end{figure}

The strength of the stochastic GW signals from the FOPT depend on $\alpha$ and $\beta$.
Larger $\alpha$ implies a stronger FOPT, while smaller $\beta$ corresponds to a longer FOPT time duration.
Consequently, larger $\alpha$ and $\tilde\beta^{-1}$ lead to stronger GW signals, as implied in Eqs.~\eqref{eq:Omega_col}, \eqref{eq:Omega_sw} and \eqref{eq:Omega_turb}.
For the surviving parameter points, we calculate the resulting values of $\alpha$ and $\tilde\beta^{-1}$, and then project the points in the $\alpha$-$\tilde\beta^{-1}$ plane, as presented in Fig.~\ref{fig:alpha_beta}.
The purple and green points are corresponding to type-I and type-II Yukawa couplings, respectively.
The parameter points lie in the ranges of $10^{-4}\lesssim \alpha \lesssim 0.3$ and $10^{-7}\lesssim \tilde{\beta}^{-1} \lesssim 0.02$.
The relic GW spectra for the parameter points are further evaluated, assuming Jouguet detonations.

We introduce $\hat\Omega_\mathrm{GW} h^2$ to denote the peak amplitudes of the GW spectra.
The contours of $\hat\Omega_\mathrm{GW} h^2$ are demonstrated in Fig.~\ref{fig:alpha_beta} and we can easily read off the GW signal strengths of the parameter points from the plot. 
The strongest GW signal we find reaches up to $\hat\Omega_\mathrm{GW} h^2 \sim 10^{-11}$.

\begin{figure}[!t]
	\centering
	\subfigure[~Type-I Yukawa couplings.\label{fig:GWScan1}]
	{\includegraphics[width=0.48\textwidth]{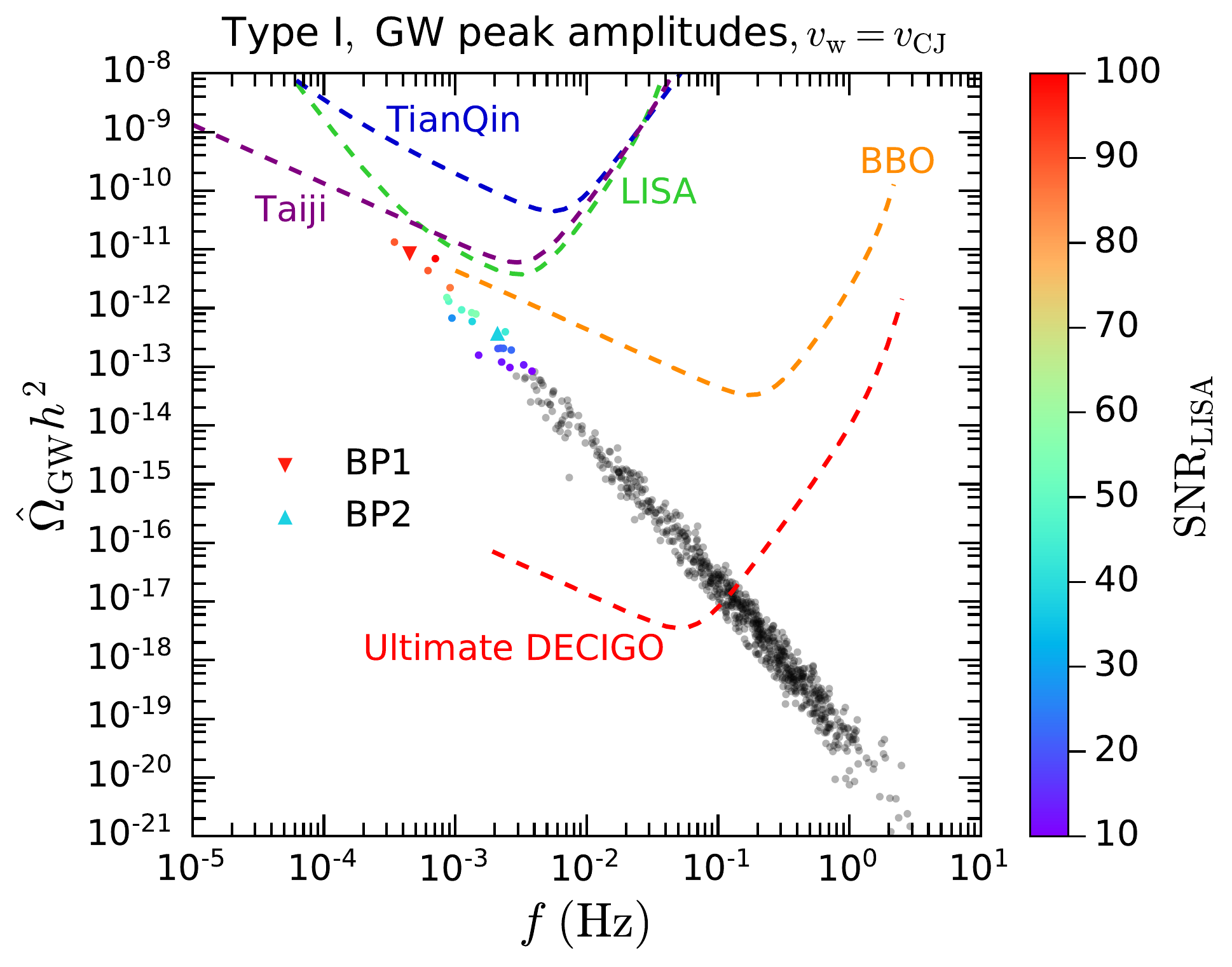}}
	\subfigure[~Type-II Yukawa couplings.\label{fig:GWScan2}]
	{\includegraphics[width=0.48\textwidth]{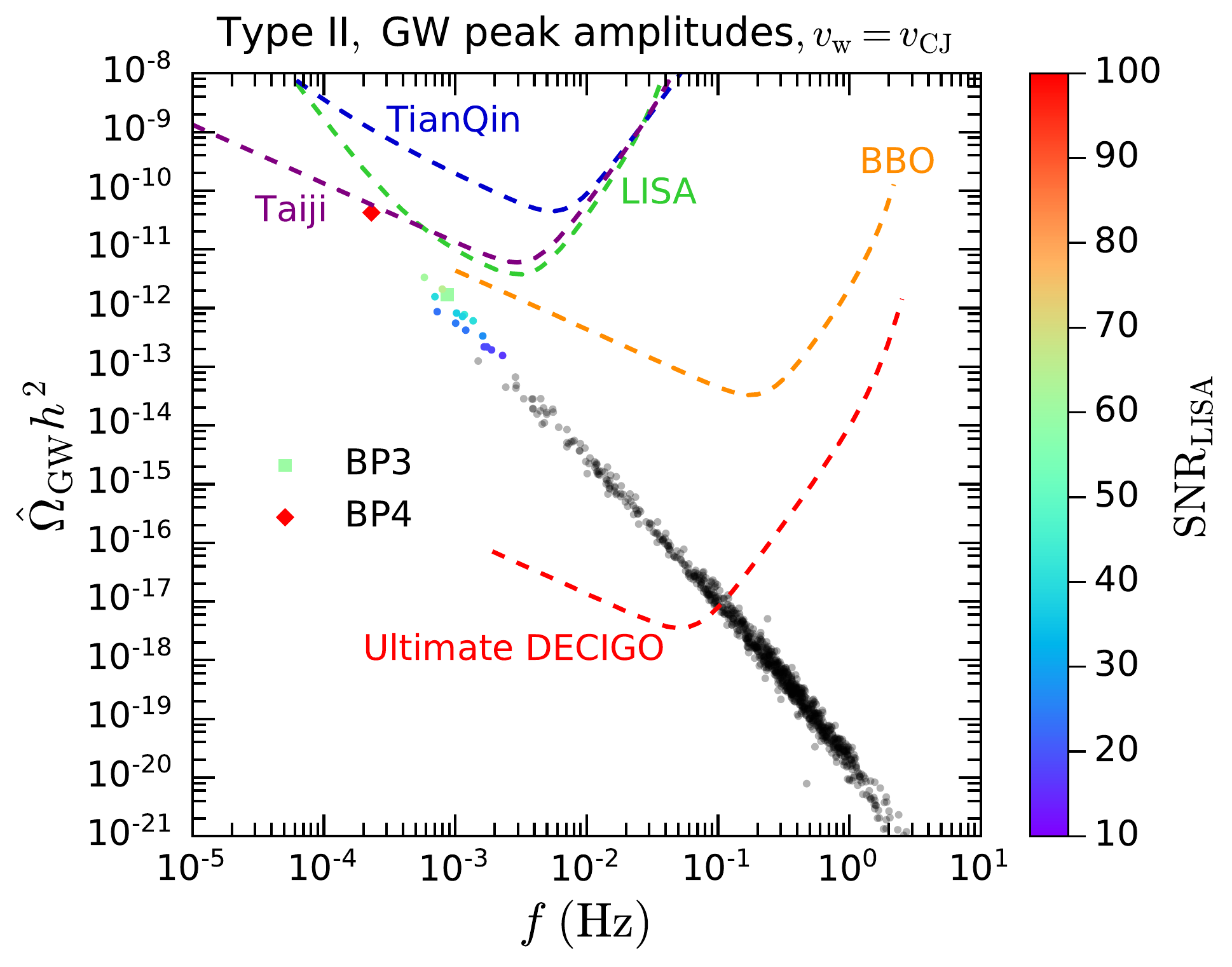}}
\caption{Peak amplitudes of the total GW spectra versus frequency for the parameter points with type-I (a) and type-II (b) Yukawa couplings assuming Jouguet detonations.
Sensitivity curves for the future space-based GW interferometers LISA~\cite{Audley:2017drz}, Tianqin~\cite{Mei:2020lrl}, Taiji~\cite{Guo:2018npi}, BBO~\cite{Cutler:2005qq}, and ultimate DECIGO~\cite{Kudoh:2005as} are also plotted. The color axes denote the LISA signal-to-noise ratio $\mathrm{SNR}_\mathrm{LISA}$ for the parameter points with $\mathrm{SNR}_\mathrm{LISA} > 10$.
The gray points yield $\mathrm{SNR}_\mathrm{LISA} < 10$.}
\label{fig:GWsScan}
\end{figure}

Fig.~\ref{fig:GWsScan} illustrates $\hat{\Omega}_\mathrm{GW}h^2$ versus the peak frequency $f$ for the parameter points.
For comparison, we also plot the sensitivity curves for the future space-based interferometers LISA~\cite{Audley:2017drz}, TianQin~\cite{Mei:2020lrl}, Taiji~\cite{Guo:2018npi}, BBO~\cite{Cutler:2005qq}, and DECIGO~\cite{Kudoh:2005as}.
Some of the curves are converted from the sensitivity on amplitude spectral density or characteristic strain.
The conversions of the related quantities can be found in, e.g., Ref.~\cite{Moore:2014lga}.
The DECIGO curve we adopt here is the ultimate sensitivity that is only limited by quantum noises, and it can be regarded as an observational limitation~\cite{Kudoh:2005as}.

The GWs produced by FOPTs become an isotropic and stochastic background in the present Universe.
The detectability of the GW signals in the  space-based interferometers increases with the practical observation time $\mathcal{T}$.
The signal-to-noise ratio can be defined as~\cite{Thrane:2013oya,Caprini:2015zlo}
\begin{equation}
\mathrm{SNR} \equiv \sqrt {\mathcal{T} \int_{f_{\min}}^{f_{\max}}{\frac{\Omega_\mathrm{GW}^2(f) }{\Omega_\mathrm{sens}^2(f)}\,df}},
\end{equation}
where $\Omega_\mathrm{sens}(f)$ is the sensitivity of the experiment.
Below we take the practical observation time $\mathcal{T} = 9.46\times 10^7~\text{s}$ (3 years) for LISA~\cite{Caprini:2019egz}, Taiji, and TianQin.
The signal $\Omega_\mathrm{GW}(f)$ can be detected if the corresponding $\mathrm{SNR}$ is larger than a signal-to-noise ratio threshold $\mathrm{SNR}_{\mathrm{thr}}$.
For the six (four) link configuration of LISA, the threshold is $\mathrm{SNR}_{\mathrm{thr}} = 10~(50)$~\cite{Caprini:2015zlo}.
We find that some parameter points yield the LISA signal-to-noise ratio $\mathrm{SNR}_\mathrm{LISA} > 10$ and could be probed by LISA.
We denote $\mathrm{SNR}_\mathrm{LISA}$ for them as the color axes in Fig.~\ref{fig:GWsScan}, with the remaining gray points corresponding to $\mathrm{SNR}_\mathrm{LISA} < 10$.
The next-generation plans aiming at $f\sim \mathcal{O}(0.1)~\si{Hz}$, like BBO and DECIGO, may probe much more parameter points.

\begin{table}[!t]
	\centering
	\setlength\tabcolsep{0.5em}
	\renewcommand{\arraystretch}{1.3}
	\caption{Detailed information for four benchmark points.}
	\label{tab:benchmarks}
	\begin{tabular}{ccccc}
		\hline
		\hline
         & BP1 & BP2 & BP3 & BP4 \\ \hline
		Type & I & I & II & II\\
		$v_s~(\mathrm{GeV})$ & $542.40$ & $384.26$ & $64.987$ & $138.82$ \\
		$m_{\chi}~(\mathrm{GeV})$ & $117.88$  & $78.191$ & $134.03$ & $76.678$ \\
		$m^2_{12}~(\mathrm{GeV}^2)$ & $2.0210\times10^4$ & $1.5876\times10^2$ & $1.7696\times10^{5}$ & $1.5042\times10^{5}$ \\
		$\tan\beta$ & $2.8616$ & $3.2654$ & $0.91655$ & $1.1732$ \\	
		$\lambda_1$ & $2.1496$ & $2.1882$ & $1.5297$ & $0.87839$ \\
		$\lambda_2$ & $0.80887$ & $0.85479$ & $1.2074$ & $0.80222$  \\
		$\lambda_3$ & $2.3925$ & $2.2628$ & $1.5741$ & $2.8002$  \\
		$\lambda_4$ & $3.0027$ & $1.4715$ & $5.3967$ & $4.4643$  \\
		$\lambda_5$ & $-6.2187$ & $-4.0567$ & $-7.8556$ & $-7.5755$  \\
		$\lambda_S$ & $3.4048$ & $2.5502$ & $6.0689$ & $4.8644$ \\
		$\kappa_1$ & $-1.4852$ & $1.0295$ & $0.80378$ & $-0.38075$ \\
		$\kappa_2$ & $1.1727$ & $-1.2142$ & $-0.83745$ & $-0.14591$ \\
		$m_{h_1}~(\mathrm{GeV})$ & $125.11$ & $91.459$ & $125.38$ & $124.87$ \\
		$m_{h_2}~(\mathrm{GeV})$ & $282.02$ & $124.77$ & $158.83$ & $307.56$ \\
		$m_{h_3}~(\mathrm{GeV})$ & $1014.5$ & $641.83$ & $650.98$ & $582.08$ \\
		$m_{a}~(\mathrm{GeV})$ & $664.75$ & $496.49$ & $911.87$ & $874.04$ \\
		$m_{H^\pm}~(\mathrm{GeV})$ & $402.96$ & $280.94$ & $655.60$ & $631.66$ \\
		$\svann_\mathrm{d}~(\mathrm{cm}^3/\mathrm{s})$ & $1.30\times10^{-26}$ & $3.68\times10^{-27}$ & $1.72\times10^{-26}$ & $6.82\times10^{-27}$\\
		$\alpha$ & $0.240$ & $0.160$ & $0.181$ & $0.346$ \\
		$\tilde{\beta}^{-1}$ & $1.33\times10^{-2}$ & $4.02\times10^{-3}$ & $7.71\times10^{-3}$ & $2.15\times10^{-2}$ \\
		$T_\mathrm{p}~(\mathrm{GeV})$ & $55.3$ & $74.9$ & $60.2$ & $47.2$ \\
		$\mathrm{SNR_{LISA}}$ & $96.6$ & $37.7$ & $60.1$ & $120$ \\
		$\mathrm{SNR_{Taiji}}$ & $83.3$ & $23.9$ & $42.3$ & $155$ \\
		$\mathrm{SNR_{TianQin}}$ & $5.50$ & $2.39$ & $3.07$ & $9.20$ \\
		\hline
		\hline
	\end{tabular}
\end{table}

For a closer look at the results, we choose four benchmark points, whose detailed information is listed in Table~\ref{tab:benchmarks}.
BP1 and BP2 (BP3 and BP4) correspond to the type-I (type-II) Yukawa couplings.
In these BPs, the masses of the Higgs bosons $h_{1,2,3}$, $a$, and $H^\pm$ are all below 1~TeV, while the mass of the DM candidate is less than 140~GeV.
The SM-like Higgs boson is $h_2$ in BP2, while it is $h_1$ in the rest BPs.
The DM annihilation cross sections $\svann_\mathrm{d}$ at dwarf galaxies predicted by the BPs are below \num{2e-26}~\si{cm^3/s}, beyond the reach of Fermi-LAT and MAGIC~\cite{Ahnen:2016qkx}.

\begin{figure}[!t]
	\centering
	{\includegraphics[width=0.55\textwidth]{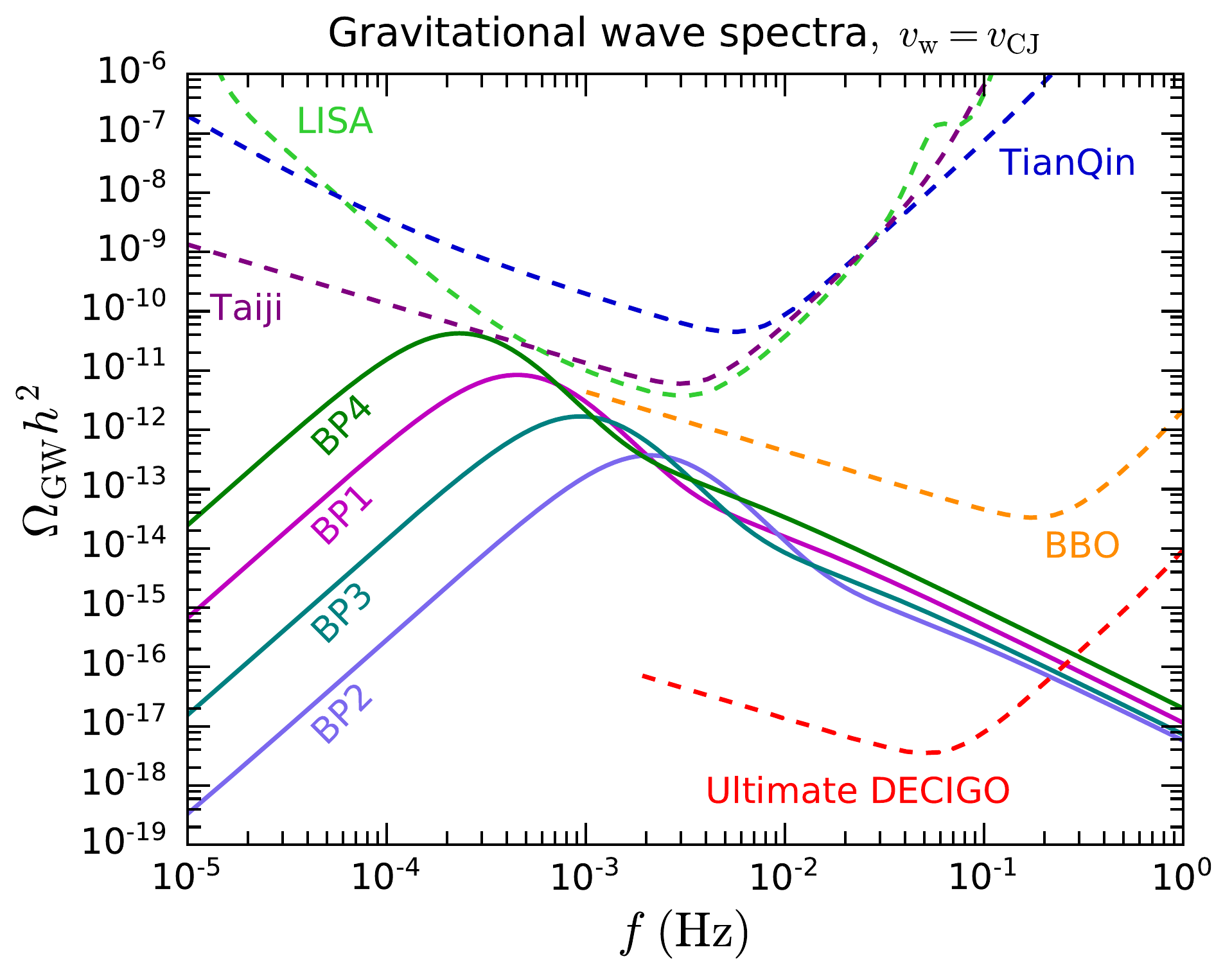}}
	\caption{GW spectra for four benchmark points assuming Jouguet detonations.}
	\label{fig:GWs}
\end{figure}

For the four BPs, percolation of the FOPT occurs in $47~\si{GeV} \lesssim T_\mathrm{p} \lesssim 75~\si{GeV}$, with $\alpha$ ranging from 0.16 to 0.35 and $\tilde\beta^{-1}$ ranging from $\num{4e-3}$ to $\num{2.2e-2}$.
Assuming Jouguet detonations, we derive the GW spectra for these BPs, as presented in Fig.~\ref{fig:GWs}.
The BPs are also indicated in Figs.~\ref{fig:alpha_beta} and \ref{fig:GWsScan}.
We find that the GW signal strengths decrease according to the order of BP4, BP1, BP3, and BP2, reflecting the descending order of $\alpha$.
In Table~\ref{tab:benchmarks}, we also list the signal-to-noise ratios $\mathrm{SNR_{LISA}}$, $\mathrm{SNR_{Taiji}}$, and $\mathrm{SNR_{TianQin}}$ for the LISA, Taiji, and TianQin experiments, respectively.
LISA and Taiji look promising to detect all BPs, while TianQin may probe BP4 with a sightly longer observation time.

\begin{figure}[!t]
	\centering
	\subfigure[~Type-I Yukawa couplings.\label{TypeIFlavorBound}]
	{\includegraphics[width=0.48\textwidth]{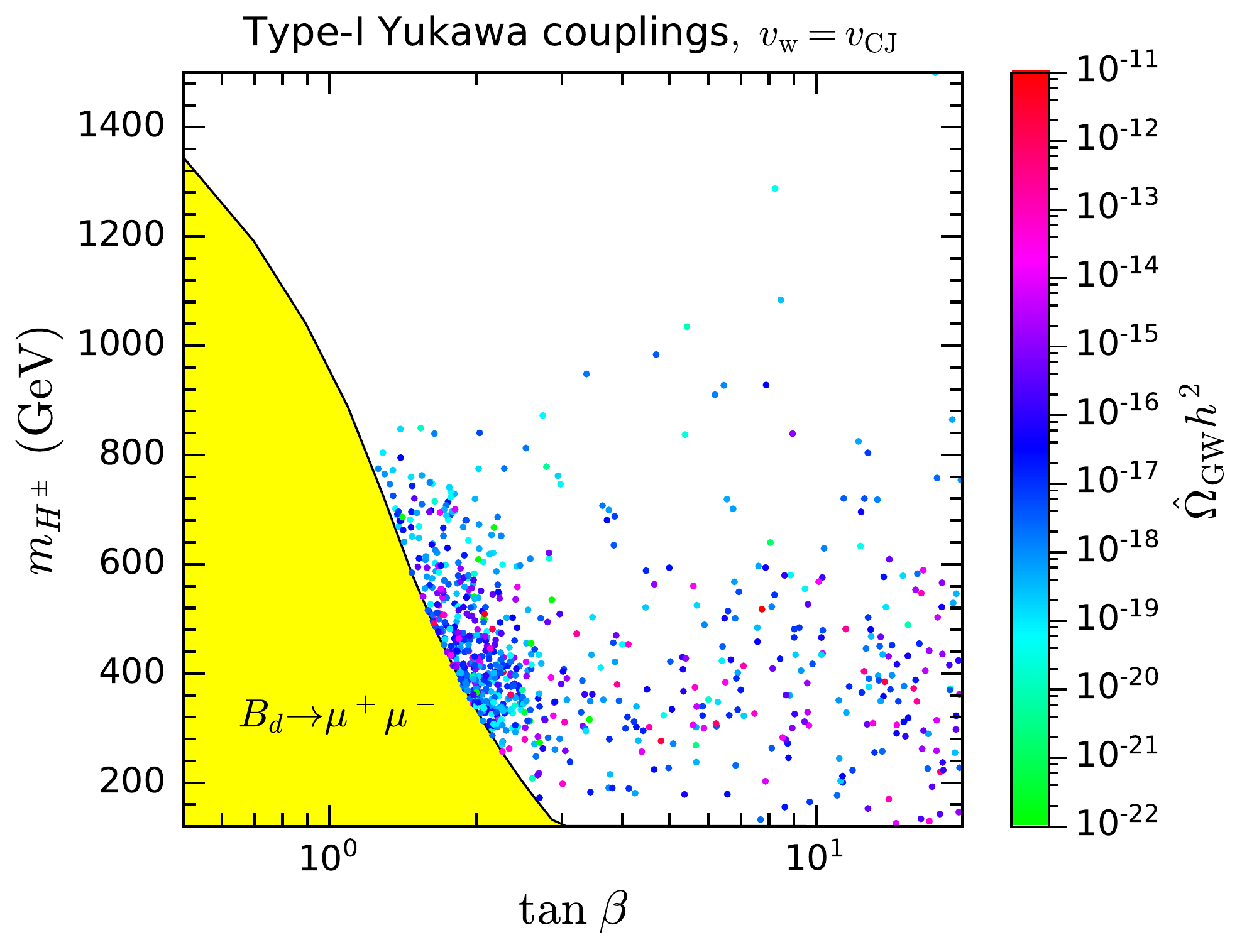}}
	\subfigure[~Type-II Yukawa couplings.\label{TypeIIFlavorBound}]
	{\includegraphics[width=0.48\textwidth]{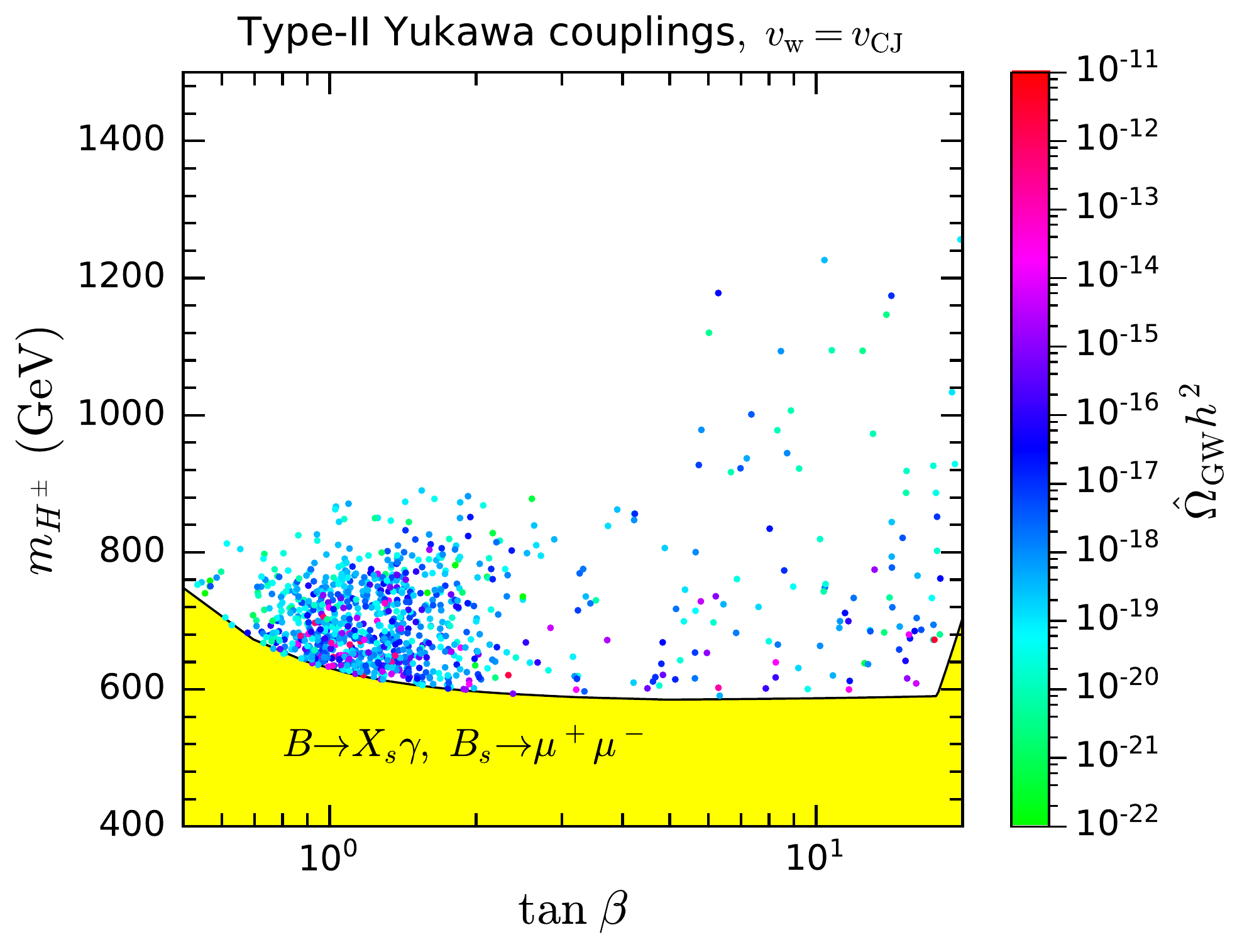}}
\caption{Parameter points projected in the $\tan\beta$-$m_H^\pm$ plane for type-I (a) and type-II (b) Yukawa couplings.
Yellow regions are excluded at 95\% C.L. by the FCNC bounds from the Gfitter global fit~\cite{Haller:2018nnx}.
The color axes denote the peak amplitude of the GW spectrum $\hat\Omega_\mathrm{GW}h^2$ for Jouguet detonations.}
\label{fig:tanbetamH+}
\end{figure}

In order to show the most important flavor constraints mentioned in Sec.~\ref{sec:bound}, we plot our parameter points confronting the FCNC bounds.
We have adopted the data from the Gfitter global fit~\cite{Haller:2018nnx} to reject parameter points.
In Fig.~\ref{fig:tanbetamH+}, the parameter points for the two types of Yukawa couplings are projected in the $\tan\beta$-$m_H^\pm$ plane, with the color axis indicating the peak amplitude of the GW spectrum, $\hat\Omega_\mathrm{GW}h^2$.

For the type-I case in Fig.~\ref{TypeIFlavorBound},
the most stringent FCNC bound comes from the LHCb and CMS measurements of $B_d \to \mu^+ \mu^-$~\cite{CMS:2014xfa,Aaij:2017vad}, excluding a region with $\tan\beta \lesssim 3$.
For the type-II case in Fig.~\ref{TypeIIFlavorBound}, the bounds from the observations of $B \to X_s \gamma$~\cite{Amhis:2016xyh,Misiak:2006zs,Misiak:2015xwa} and
$B_s \to \mu^+ \mu^-$~\cite{CMS:2014xfa,Aaij:2017vad} exclude a region with $m_{H^\pm} \lesssim 750~\si{GeV}$.
Thus, the FCNC constraints remove small $\tan\beta$ and light $H^\pm$ for type-I and type-II Yukawa couplings, respectively.
We remark that strong GW signals typically favor small $m_{H^\pm}$.
There are many parameter points with $m_{H^\pm}\lesssim 600~\si{GeV}$ in the type-I case leading to large $\hat\Omega_\mathrm{GW}h^2$.
In the type-II case, the $m_{H^\pm}\lesssim 600~\si{GeV}$ region is basically excluded by the FCNC constraints, and hence it is more difficult to achieve strong GW signals.

\begin{figure}[!t]
	\centering
	\subfigure[~Type-I Yukawa couplings.\label{TypeI_mchi_svd}]
	{\includegraphics[width=0.48\textwidth]{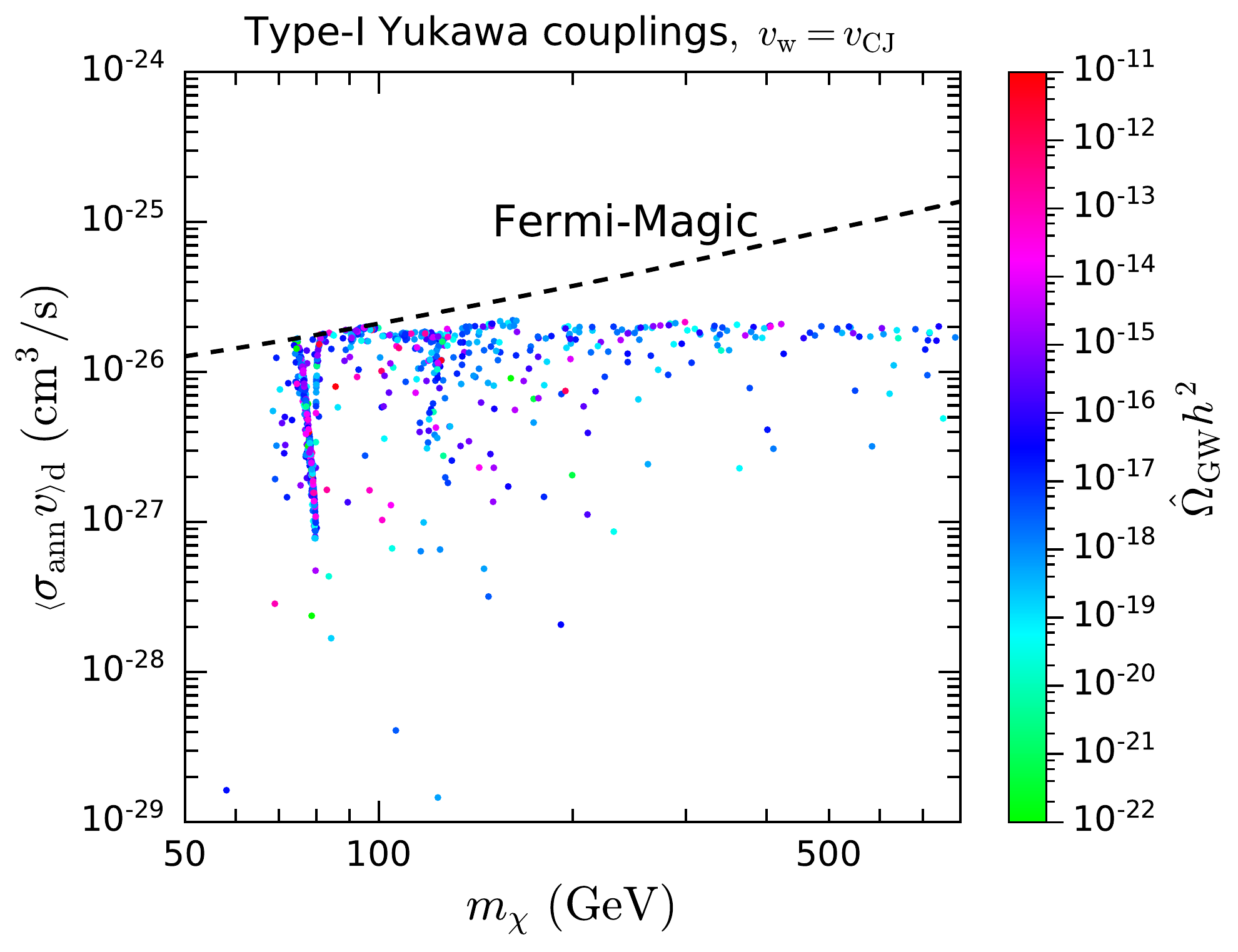}}
	\subfigure[~Type-II Yukawa couplings.\label{TypeII_mchi_svd}]
	{\includegraphics[width=0.48\textwidth]{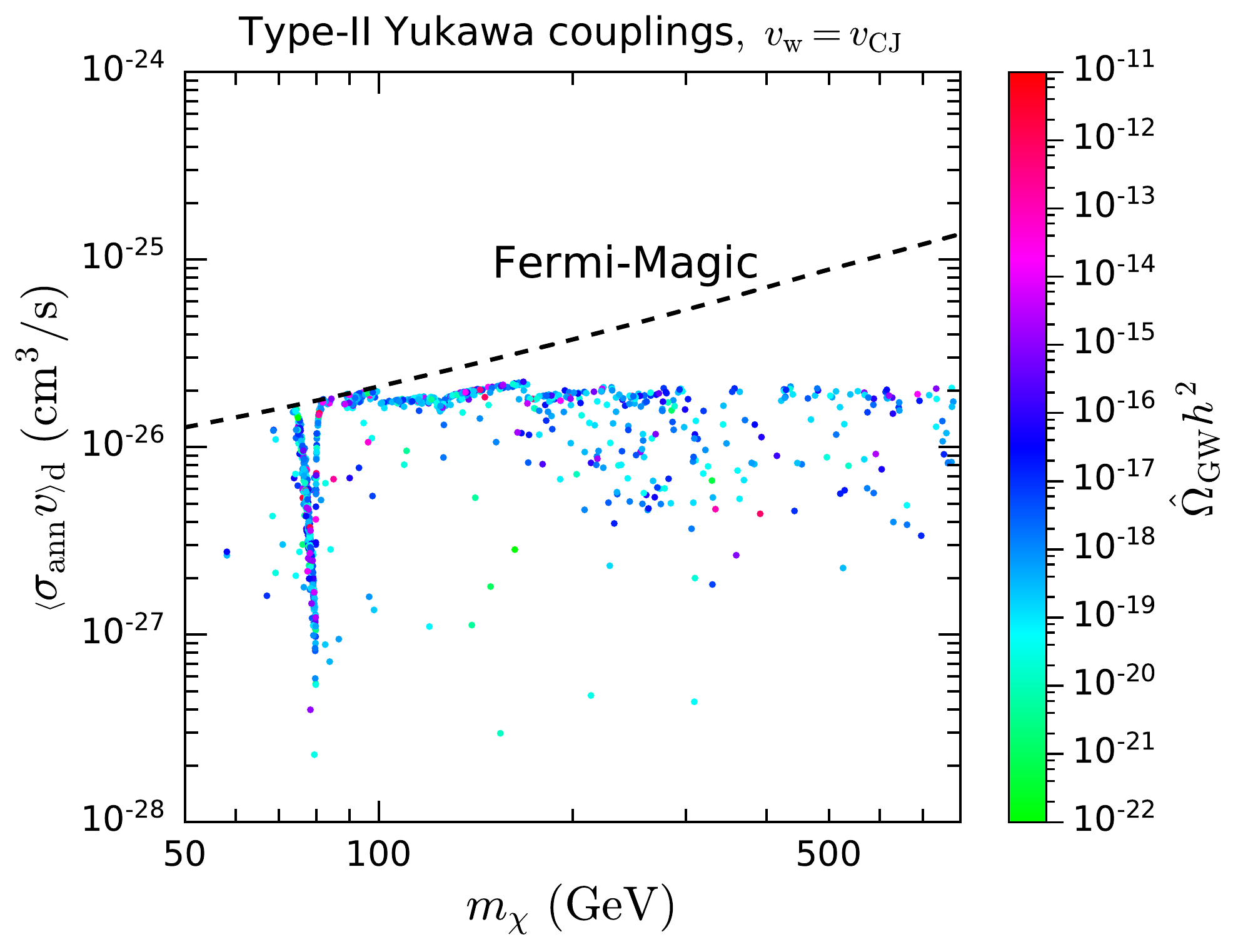}}
\caption{Parameter points projected in the $m_{\chi}$-$\svann_\mathrm{d}$ plane for type-I (a) and type-II (b) Yukawa couplings.
Dashed lines denote the 95\% C.L. upper limits on the DM annihilation cross section from the $\gamma$-ray observations of dwarf galaxies by Fermi-LAT and MAGIC~\cite{Ahnen:2016qkx}.}
\label{fig:mchi_svd}
\end{figure}

In Fig.~\ref{fig:mchi_svd}, we project the parameter points in the $m_{\chi}$-$\svann_\mathrm{d}$ plane.
Although all these parameter points are required to predict the observed relic abundance, $\svann_\mathrm{d}$ can deviate from the canonical annihilation cross section $3\times 10^{-26}~\si{cm^3/s}$, due to the velocity dependence of $\svann$.
One reason is that some annihilation channels kinematically forbidden at low velocities could be opened at the freeze-out epoch, and another reason is the resonance effects~\cite{Griest:1990kh}.
The pile-up of points around $m_\chi \sim 78~\si{GeV}$ is mainly related to the annihilation channel $\chi\chi\to W^+W^-$.

In the above numerical analyses, we have assumed the bubble propagation mode to be Jouguet detonations with bubble wall velocity $v_\mathrm{w} = v_\mathrm{CJ}$.
Below, we study the effect of various bubble propagation modes.
In general, the dependence of $\kappa_v$  on $v_\mathrm{w}$ and $\alpha$ can be found in Ref.~\cite{Espinosa:2010hh}.
The sound speed $c_\mathrm{s}$ in the relativistic plasma is very close to $1/\sqrt{3}$.
For $v_\mathrm{w} \ll c_\mathrm{s}$, $v_\mathrm{w} = c_\mathrm{s}$, and $v_\mathrm{w} = 1$, the vacuum energy fraction converted into the bulk kinetic energy of the fluid $\kappa_v$ has the following analytic approximations, based on fit.
\begin{align}
v_\mathrm{w} \ll c_\mathrm{s}: &\quad \kappa _v^{\mathrm{A}} = \frac{{6.9\alpha v_{\mathrm{w}}^{6/5}}}{{1.36 - 0.037\sqrt \alpha   + \alpha }}.\\
v_\mathrm{w} = c_\mathrm{s}: &\quad \kappa _v^{\mathrm{B}} = \frac{{{\alpha ^{2/5}}}}{{0.017 + {(0.997 + \alpha )^{2/5}}}}.\\
v_\mathrm{w} = 1: &\quad \kappa _v^{\mathrm{D}} = \frac{\alpha }{{0.73 + 0.083\sqrt \alpha   + \alpha }}.
\end{align}
Furthermore, for subsonic deflagrations (${v_{\mathrm{w}}} < {c_{\mathrm{s}}}$), supersonic deflagrations (${c_{\mathrm{s}}} < {v_{\mathrm{w}}} < {v_{{\mathrm{CJ}}}}$), and detonations (${v_{\mathrm{w}}} \gtrsim {v_{{\mathrm{CJ}}}}$), $\kappa_v$ is roughly given by
\begin{eqnarray}
{\kappa _v}({v_{\mathrm{w}}} < {c_{\mathrm{s}}}) &=& \frac{{c_{\mathrm{s}}^{11/5}\kappa _v^{\mathrm{A}}\kappa _v^{\mathrm{B}}}}{{(c_{\mathrm{s}}^{11/5} - v_{\mathrm{w}}^{11/5})\kappa _v^{\mathrm{B}} + {v_{\mathrm{w}}}c_{\mathrm{s}}^{6/5}\kappa _v^{\mathrm{A}}}},
\\
{\kappa _v}({c_{\mathrm{s}}} < {v_{\mathrm{w}}} < {v_{{\mathrm{CJ}}}}) &=& \kappa _v^{\mathrm{B}} + ({v_{\mathrm{w}}} - {c_{\mathrm{s}}})\delta \kappa  + {\left( {\frac{{{v_{\mathrm{w}}} - {c_{\mathrm{s}}}}}{{{v_{{\mathrm{CJ}}}} - {c_{\mathrm{s}}}}}} \right)^3}\left[\kappa _v^{\mathrm{CJ}} - \kappa _v^{\mathrm{B}} - ({v_{{\mathrm{CJ}}}} - {c_{\mathrm{s}}})\delta \kappa \right],
\\
{\kappa _v}({v_{\mathrm{w}}} \gtrsim {v_{{\mathrm{CJ}}}}) &=& \frac{{{{({v_{{\mathrm{CJ}}}} - 1)}^3}{{({v_{{\mathrm{CJ}}}}/{v_{\mathrm{w}}})}^{5/2}}\kappa _v^{\mathrm{CJ}}\kappa _v^{\mathrm{D}}}}{{[{{({v_{{\mathrm{CJ}}}} - 1)}^3} - {{({v_{\mathrm{w}}} - 1)}^3}]v_{{\mathrm{CJ}}}^{5/2}\kappa _v^{\mathrm{CJ}} + {{({v_{\mathrm{w}}} - 1)}^3}\kappa _v^{\mathrm{D}}}},
\end{eqnarray}
where $\delta \kappa  =  - 0.9\ln [\sqrt \alpha/(1 + \sqrt \alpha)]$.

\begin{figure}[!t]
	\centering
	{\includegraphics[width=0.55\textwidth]{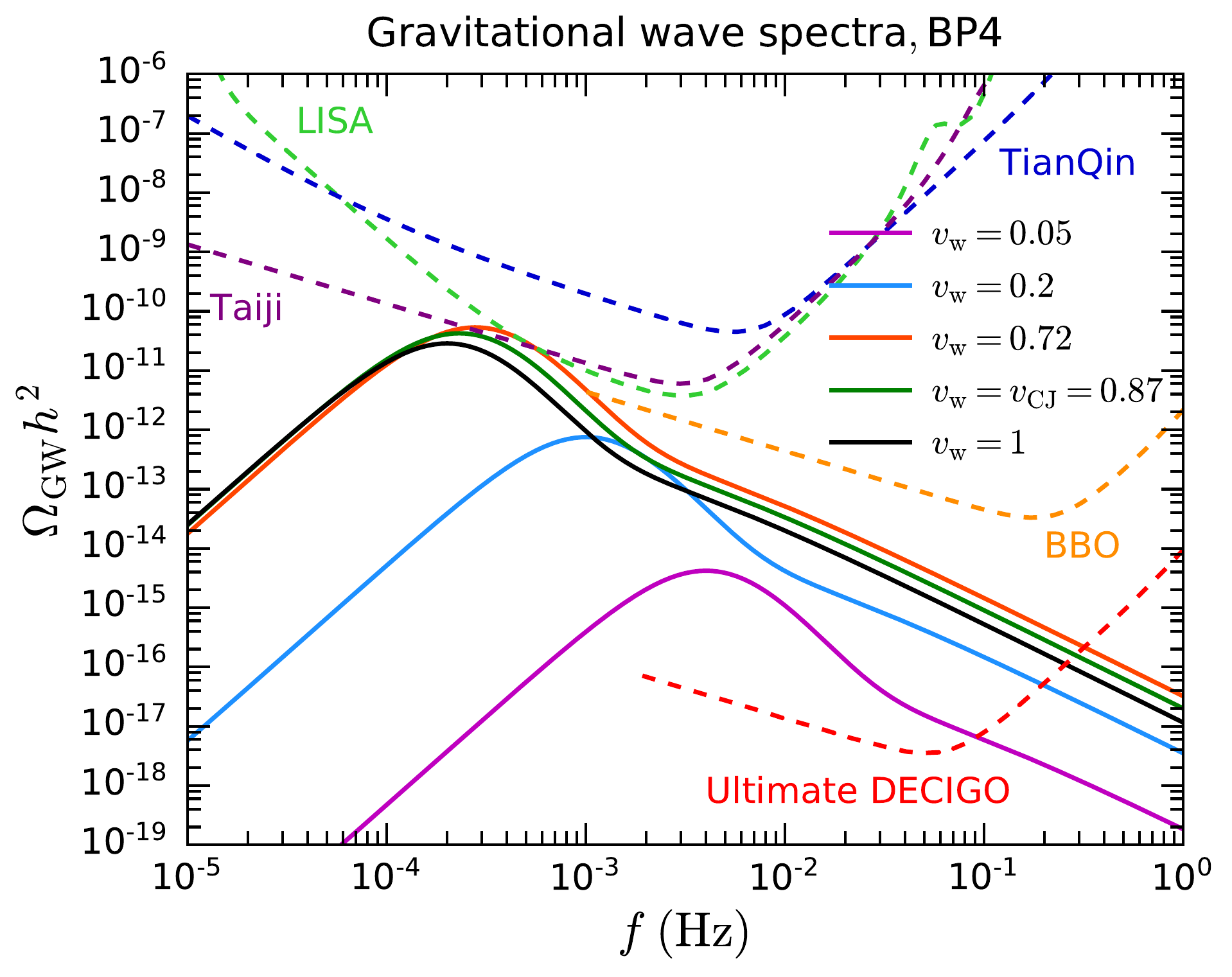}}
	\caption{GW spectra for BP4 with several assumptions of $v_\mathrm{w}$.}
	\label{fig:GW_vw}
\end{figure}

According to these expressions, we derive the GW spectra for BP4 assuming the bubble wall velocity $v_\mathrm{w} = 0.05$, $0.2$, $0.72$, and $1$.
These spectra are demonstrated in Fig.~\ref{fig:GW_vw}, along with the previously obtained BP4 spectrum for $v_\mathrm{w}= v_\mathrm{CJ} = 0.87$.
Compared to Jouguet detonations with $v_\mathrm{w} = v_\mathrm{CJ}$, supersonic deflagrations with $v_\mathrm{w} = 0.72$ lead to a stronger GW signal, which could be properly tested by TianQin with $\mathrm{SNR}_{\mathrm{TianQin}} = 15.8$.
On the other hand, subsonic deflagrations with $v_\mathrm{w} = 0.05$ and $0.2$ give much weaker GW signals.

\section{Summary}
\label{sec:concl}

In this paper, we have studied the stochastic GW signals from electroweak FOPTs in the model comprising the pNGB dark matter framework and two Higgs doublets with type-I or type-II Yukawa couplings.
The DM candidate is a pNGB whose tree-level scattering off nucleons vanishes at zero momentum transfer, evading the constraints from direct detection experiments.
The three scalar fields in the model have nonzero VEVs at zero temperature, which should be developed from EWPTs at the early Universe.
If such EWPTs are of strongly first order, stochastic GWs could be effectively produced.
The effective potential has been carefully constructed with one-loop corrections at zero temperature as well as thermal corrections, allowing us to carry out accurate analyses on EWPTs.

We have performed random scans in the 12-dimensional parameter space, taking into account the constraints from bounded from below conditions, LHC run~1 and run~2 measurements of the 125~GeV Higgs boson, FCNC $B$-meson decays, the  Planck observation of the DM relic abundance, and the $\gamma$-ray observations of dwarf galaxies by Fermi-LAT and MAGIC.
The surviving parameter points are also required to  induce an electroweak FOPT.
We have further analyzed the characteristic temperatures, the phase transition strength, and the characteristic time duration of the FOPT.
Based on such information, the resulting relic GW spectra from sound waves and MHD turbulence have been evaluated.

Assuming that the bubble propagation mode is Jouguet detonations with $v_\mathrm{w} = v_\mathrm{CJ}$, we have found that the FOPTs of some parameter points could induce peak amplitudes of the GW spectra around $10^{-13} \text{--} 10^{-11}$, which could be well detected by the future space-based GW interferometers LISA and Taiji.
The next-generation GW interferometers BBO and DECIGO are capable of probing much more parameter points.
We have noticed that a lighter charged Higgs boson $H^\pm$ in this model is more probable to induce a strong GW signal.
Since the FCNC constraints on $m_{H^\pm}$ for type-II Yukawa couplings at large $\tan\beta$ are more stringent than those for type-I Yukawa couplings, the type-I case typically leads to stronger GW signals.

We have also investigated the effects of different bubble propagation modes with several values of the bubble wall velocity $v_\mathrm{w}$.
For the benchmark point BP4, supersonic deflagrations with $v_\mathrm{w} = 0.72$ can induce a stronger GW signal than Jouguet detonations with $v_\mathrm{w} = v_\mathrm{CJ}$.
In this optimistic case, BP4 could be well tested by LISA, Taiji, and TianQin.
Detonations with $v_\mathrm{w} = 1$ lead to a slightly weaker GW signal, while subsonic deflagrations with $v_\mathrm{w} = 0.2$ and $0.05$ result in much weaker signals.

\begin{acknowledgments}

We thank Fa Peng Huang and Ligong Bian for helpful discussions. This work is supported in part by the National Natural Science Foundation of China under Grants No.~11805288, No.~11875327, No.~11905300, and No.~12005312, the China Postdoctoral Science Foundation
under Grant No.~2018M643282,
the Fundamental Research Funds for the Central Universities,
and the Sun Yat-Sen University Science Foundation.

\end{acknowledgments}

\bibliographystyle{utphys}
\bibliography{ref}

\end{document}